\documentclass[fleqn,usenatbib]{mnras}

\usepackage{newtxtext,newtxmath}

\usepackage[T1]{fontenc}

\DeclareRobustCommand{\VAN}[3]{#2}
\let\VANthebibliography\thebibliography
\def\thebibliography{\DeclareRobustCommand{\VAN}[3]{##3}\VANthebibliography}

\usepackage{graphicx}	
\usepackage{amsmath}	
\graphicspath{{images/}}
\usepackage{multicol}
\usepackage{caption}
\usepackage{xcolor}
\usepackage{url}
\usepackage{placeins}
\usepackage{rotating}
\usepackage{siunitx}
\usepackage{multirow}
\usepackage{booktabs}
\usepackage{floatrow}
\usepackage{subcaption}

\DeclareMathOperator{\sign}{sgn}

\title[Geomagnetic storms and coronal holes]{
\textbf{Geomagnetic storm forecasting from solar coronal holes}}        

\author[S. Nitti et al.]{
Simona Nitti,$^{1}$\thanks{E-mail: simona.nitti@skoltech.ru}
Tatiana Podladchikova,$^{1}$
Stefan J. Hofmeister,$^{2}$
Astrid M. Veronig,$^{3,4}$\newauthor
Giuliana Verbanac$^{5}$ and
Mario Bandić $^{6}$
\\
$^{1}$Skolkovo Institute of Science and Technology,
Bolshoy Boulevard 30, bld. 1,
121205, Moscow, Russia\\
$^{2}$Leibniz Institute for Astrophysics Potsdam, An der Sternwarte 16, 14482, Potsdam, Germany\\
$^{3}$Institute of Physics, University of Graz, 	Universit{\"a}tsplatz 5, 
A-8010, Graz, Austria\\
$^{4}$Kanzelh{\"o}he Observatory for Solar and Environmental Research, University of Graz, Kanzelh{\"o}he 19, 
9521, Treffen, Austria\\
$^{5}$Department of Geophysics, Faculty of Science, University of Zagreb, Horvatovac 95, 10000, Zagreb, Croatia\\
$^{6}$Astronomical Observatory Zagreb, Opatička 22, 10000, Zagreb, Croatia
}

\date{Accepted 2022 November 28. Received 2022 November 17; in original form 2022 June 14}

\pubyear{2022}

\begin{document}
\label{firstpage}
\pagerange{\pageref{firstpage}--\pageref{lastpage}}
\maketitle

\begin{abstract}
Coronal holes (CHs) are the source of high-speed streams (HSSs) in the solar wind, whose interaction with the slow solar wind creates corotating interaction regions (CIRs) in the heliosphere. Whenever the CIRs hit the Earth, they can cause geomagnetic storms. We develop a method to predict the strength of CIR/HSS-driven geomagnetic storms directly from solar observations using the CH areas and associated magnetic field polarity. First, we build a dataset comprising the properties of CHs on the Sun, the associated HSSs, CIRs, and orientation of the interplanetary magnetic field (IMF) at L1, and the strength of the associated geomagnetic storms by the geomagnetic indices Dst and Kp. Then, we predict the Dst and Kp indices using a Gaussian Process model, which accounts for the annual variation of the orientation of Earth's magnetic field axis.
We demonstrate that the polarity of the IMF at L1 associated with CIRs is preserved in around 83\% of cases when compared to the polarity of their CH sources. Testing our model over the period 2010-2020, we obtained a correlation coefficient between the predicted and observed Dst index of $\text{R} = 0.63/0.73$, and Kp index of $\text{R} = 0.65/0.67$, for HSSs having a polarity towards/away from the Sun. These findings demonstrate the possibility of predicting CIR/HSS-driven geomagnetic storms directly from solar observations and extending the forecasting lead time up to several days, which is relevant for enhancing space weather predictions.
\end{abstract}




\section{Introduction}

Coronal holes (CHs) are large low density and low temperature regions in the solar corona. Therefore, they appear as dark regions when measured in extreme ultraviolet (EUV) and X-ray radiation. CHs exhibit a dominant magnetic field polarity which results in an 'open' magnetic field topology, i.e., the magnetic field reaches far into interplanetary space before it closes back to the Sun. 
Along these field lines, solar plasma is accelerated towards interplanetary space, forming the so-called high-speed solar wind streams (HSSs) with velocities up to 800 km/s \citep[e.g., review by][]{Cranmer2009}. The magnetic field lines and HSSs are frozen-in together from their source on the Sun, and travel away from the rotating Sun towards interplanetary space forming an Archimedian spiral known as the Parker spiral \citep{Parker1958}. 
Along the spiral trajectory, HSSs collide into the ambient slow solar wind, causing a compression of both plasma and magnetic fields on the rising-speed portion of the HSS \citep[e.g. ][]{Parker1963,Sarabhai1963,Carovillano1969,Gosling1972}. The pattern of compression rotates with the Sun, and gives these regions the name of corotating interaction regions \citep[CIRs;][]{Smith1976}.
 
The propagation time of a CIR from the Sun to Earth is about 2--6 days.
When a CIR hits the Earth, the interplanetary magnetic field (IMF) carried by the CIR can interact with the Earth's magnetosphere, enabling a magnetic reconnection. A magnetic reconnection, being the trigger of a geomagnetic storm, is a recombination in the magnetic field topology with a sudden release of solar wind mass, energy and momentum towards Earth. It occurs when the IMF points southward, i.e., it is antiparallel to the Earth's magnetic field. This corresponds to a negative Bz component of the IMF in GSM coordinates, usually expressed by Bs, i.e., Bs = Bz when Bz < 0 and Bs is not defined otherwise \citep[see for example ][]{Verbanac2021}. 
As Earth rotates around the Sun, the orientation of Earth's magnetic field axis varies over the year. When we assume the IMF lying along an Archimedian spiral, the varying angle between Earth's magnetic field axis and the IMF varies the chance for magnetic reconnection over the year. This is known as the Russel McPherron (RM) effect \citep{Russell1973}. For polarities towards the Sun it results in the largest Bs in spring equinox, and for IMF polarities directed away from the Sun in fall equinox.
A shortcoming of the RM model is that the IMF is assumed to be in the direction of an ideal Parker spiral whereas the IMF vector varies around this ideal direction due to alfvenic oscillations and turbulences.
From the in situ IMF data, \citet{Verbanac2021} have recently obtained the statistical patterns of Bs and Bs ordered according to the IMF polarity through a solar cycle. 
They showed that Bs ordered according to the polarity differs from the one predicted by the RM model and exhibits a \textit{pair of spectacles} pattern: when the IMF points toward/away from the Sun the field is enhanced in spring/fall and reduced in fall/spring. Therefore, the pattern of an experimentally measured Bs ordered according to the polarity differs from those predicted by the RM model, and the differences are due to the neglection of the IMF variations around the ideal direction used in the RM model. 
The product of Bs with the solar wind velocity, Bsv, plays a crucial role for the magnetospheric dynamics, and drives geomagnetic activity \citep{Prolss2004}. The geomagnetic activity is measured, amongst others, by the Dst index and by the Kp index. 
The Dst index is associated with the geomagnetic activity at low latitudes caused by ring current and tail currents, and is measured by the depression of the equatorial magnetic field at the Earth's surface. The Kp index is quasi-logarithmic index, associated to the geomagnetic activity at mid latitudes \citep{Prolss2004}.

The majority of Dst and Kp forecasts at Earth employ the in situ IMF and solar wind observations in the libration point L1. There are three main approaches: 
(1) empirical approaches that analyze the relations between Dst peaks and geoefficient solar wind and IMF parameters
\citep[e.g.][]{Akasofu1981, Petrukovich2000, Gonzalez2005, Kane2007, Mansilla2008, Yermolaev2010, Echer2013, Rathore2014}; 
(2) analytical approaches based on first-order differential equations \citep[e.g.][]{Burton1975,OBrien2000a,OBrien2000b,Podladchikova2012storms,Podladchikova2018storms} and  magnetospheric models \citep{Katus2015, Rastatter2013} to describe the Dst evolution as a function of geoeffective solar wind parameters; 
(3) machine learning approaches to predict the Dst and Kp index, that use statistical models, artificial neural networks and nonlinear auto-regression techniques \citep[e.g. ][]{Lundstedt2002, Temerin2002, Wei2004, Wing2005,Pallocchia2006, Sharifie2006, Zhu2006, Amata2008, Bala2012,Revallo2014, Andriyas2015,Shprits2019}.
As the solar wind parameters are measured only close to Earth, the lead warning time of these models is only 1-2 hours. 

To increase the lead warning time to several days, solar data from the vicinity of the Sun has to be employed. One could either forecast the solar wind properties near Earth from solar observations or the geomagnetic effects directly from solar observations. The former has been done, e.g., by using an empirical relationship between the area of coronal holes and the solar wind velocity near Earth \citep[e.g.,][]{Nolte1976, Robbins2006, Vrsnak2007a, Verbanac2011a, Verbanac2011b, Verbanac2013, Rotter2012, Rotter2015, Hofmeister2018}. This relationship was recently explained by a propagational effect of HSSs from the Sun to Earth \citep{Hofmeister2020, Hofmeister2022}. The latter has been done by \cite{Vrsnak2007b}. They focused on 100 days in 2005 when the solar wind was modulated primarily by HSSs, producing a relationship between CH areas and the Dst index. \cite{Verbanac2011a} focused on the same study period, but beside the Dst-CH relationship additionally provided a relationship between CH area and the Ap index, a linearly scaled Kp index. Both studies accounted for the annual variation of Bs, by using the CH polarity. They showed that the geomagnetic indices ordered by the polarity contains the imprint of Bs which can be employed to mimic the variations in geomagnetic indices, notwithstanding the physics that explains it.

In the present study, we analyze the relationships of geomagnetic indices Dst and Kp separately for coronal holes having a polarity toward and away from the Sun, during the decade 2010--2020. This allows us to relate the polarity of the coronal holes to the strength of geomagnetic storms over the year and to analyze this relationship in detail. From this analysis, we provide a novel forecasting model for geomagnetic indices Dst and Kp directly from solar observations. Since the propagation time of solar wind from the Sun to Earth is about 2--6 days, predicting geomagnetic storms from their sources on the Sun increases the warning times from hours to days.

\section{Data and data preparation}\label{Datasets_Data_Reduction}
\begin{figure}
   \centering
    {\includegraphics[width=0.88\columnwidth]{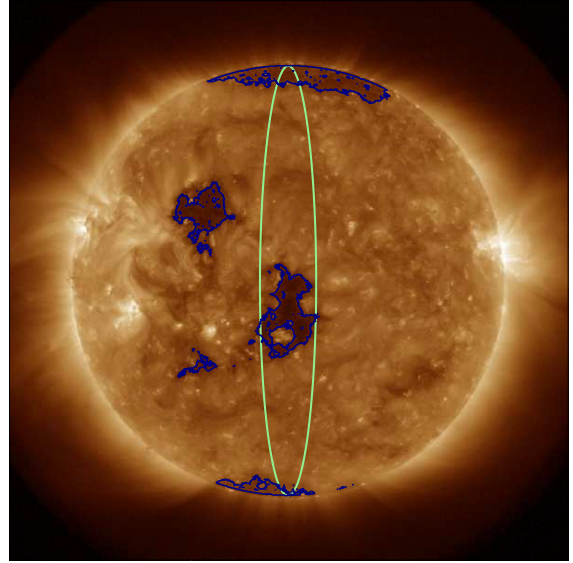}}
    \caption{ SDO/AIA 193~Å filtergram recorded on 2017 June 13 at 18:00:04 UT. The blue-outlined structures are the CHs on the solar corona. The green ellipse is the 15 degrees wide central meridional slice. }
    \label{fig:CHmask}
\end{figure}
We use the following data sets: CH fractional area and mean magnetic flux density within the CH, in situ solar wind velocity and IMF strength, and geomagnetic indices Dst and Kp.
We analyze the period from October 2010 to December 2020, covering most of solar cycle 24, and we utilize time series at a cadence of 1 hour. 

We employ the operational tool hosted at the University of Graz to detect coronal hole regions. It analyzes EUV 193~Å filtergrams, recording the solar corona at a temperature of 1.6~MK taken by Atmospheric Imaging Assembly \cite[AIA;][]{Lemen2012} onboard the Solar Dynamics Observatory \cite[SDO;][]{Pesnell2012}. Then, a segmentation method based on the intensity distribution is applied to identify CHs regions \cite[for more details, see][]{Rotter2012}. We define the CH area as the fractional area CHs cover within a 15 degrees wide meridional slice around the solar central meridian. Figure \ref{fig:CHmask} shows the central meridional slice as seen from Earth in the helioprojective coordinate frame, and the CHs detected by the segmentation method. 
The segmented CH masks within the central meridional slice are projected and co-registered to the photospheric line-of-sight magnetograms measured by the Helioseismic Magnetic Imager \cite[HMI;][]{Scherrer2012} onboard SDO. They provide an estimate of the mean magnetic flux density of the CHs and their polarities \cite[cf.,][]{Hofmeister2017}. 
For the solar wind speed and the IMF, we use data measured in situ at L1 and retrieved via the OMNI database.
For the solar wind speed, it uses measurements from the Solar Wind Electron, Proton and Alpha Particle Monitor \cite[SWEPAM;][]{McComas1998} onboard Advanced Composition Explorer \cite[ACE;][]{Stone1998} and Solar Wind Experiment \cite[SWE;][]{Ogilvie1995} onboard the Wind spacecraft \citep{Ogilvie1997}, propagated forward in time to the Earth bow shock. The IMF is measured by input data from the ACE’s Magnetometer \citep[MAG;][]{Smith1998}) and the Wind’s Magnetic Field Investigation \citep[MFI;][]{Lepping1995} measurements.
The Dst index is retrieved from the World Data Center (WDC) for Geomagnetism in Kyoto\footnote{\url{https://wdc.kugi.kyoto-u.ac.jp/dstdir}} . 
Dst is derived from hourly values of horizontal magnetic field variations measured at four low-latitude geomagnetic observatories between 28° and 34° northern or southern geomagnetic latitude.
The Kp index is calculated by the German GeoForschungsZentrum (GFZ) in Potsdam
\footnote{\url{https://datapub.gfz-potsdam.de/download/10.5880.Kp.0001/Kp_definitive/}}. 
Kp is the arithmetic mean of the 3-hour quasi-logarithmic K-index from 13 observatories between 44° and 60° northern or southern geomagnetic latitude.
To reduce the statistical noise, we smoothed all the data sets with a forward-backward exponential smoothing method \citep{Brown1963}.  It uses an exponential window function with a single parameter, called smoothing factor, to reduce the response to rapid changes (such as noise) in the input signal. It is set to a value between 0 and 1, where lower values cause the output to respond slowly to changes and vice-versa. In our analysis, the smoothing factor is set equal to 0.40.

\subsection{Gaussian Process Regression}
In this study, we use the Gaussian Process regression \citep[GPR;][]{Rasmussen2005} to fit the functions of solar wind speed, v, and the ratios Dst/v and Kp/v. For more information, see Appendix \ref{Sect:GPR}.
The GPR can conveniently be applied to fit very flexible non-linear problems when the number of available data is limited, as in the present case. 
It assumes the observations $y$ approximated at input $x$ by a predictive function $f$, from an ensemble of functions. 
The function $f$ is a Gaussian process (GP) specified by its mean and standard deviation functions, written as $f(x) \sim GP(m(x),\,\sigma(x))$. In this study, we consider as predictive function only the mean $m(x)$ of $f$. This corresponds to the average of the ensemble of realizations.
To get the formulae of $m(x)$ and $\sigma(x)$, the GPR requires the choice of a prior function, also called kernel, which bears the information about the shape and structure the output $y$ is expected to have. The shape of the kernel function is modulated by the hyperparameters $\sigma_K$ and $l_K$. Supposing our output $y$ has a non-null noise in its measurement, we add a third hyperparameter, $\sigma_\textrm{n}$. $\sigma_\textrm{n}$ is an independent identically distributed Gaussian noise that measures the uncertainty in our output.
Finding the optimal hyperparameters, i.e the set of hyperparameters that best approximate the problem, is an optimization problem. The objective function to be optimized is the negative log-likelihood, described in Equation~\ref{eq:loss function}.
Substituting the hyperparameters in the formulae of $m(x)$ and $\sigma(x)$, the predictive function $f$ is completely defined. 

\section{Forecasting model} \label{forecasting}
The geomagnetic Dst and Kp indices depend on Bs and the solar wind velocity. 
\begin{equation}
Dst \sim Bs\cdot v.
\end{equation}
When building the model of Dst and Kp, we try to take their effects into account separately. The Bs effect is linked to the polarity of the IMF, P$_\textrm{L1}$, and the orientation of the Earth’s magnetic field axis, that depends on the day of the year (DOY). In turn, the polarity of the IMF and the orientation of the Earth's magnetic field axis impact the Dst index normalized to the solar wind velocity, 
\begin{equation}
 Bs \sim f_1(P_\textrm{L1}, DOY) \sim Dst/v.
\end{equation}
The polarity of the IMF in CIRs at L1 is related to the polarity of the CH sources on the Sun as their solar source regions \citep{Choi2009}. 
Hence, we use the polarity of the CHs in the place of the IMF polarity at L1 to rely only on solar observations.
Then, we account for the effect of the solar wind velocity by studying its relationship with the area of CHs within the central meridional slice,  $v \sim f_2(A_\textrm{CH})$. Combining these effects, we are able to predict Dst and Kp as:
\begin{equation}
Dst = (Dst/v) \cdot v \sim f_1(P_\textrm{L1}, DOY) \cdot f_2(A_\textrm{CH}).
\end{equation}
The development of the forecasting model comprises 6 steps:
\begin{enumerate}
    \item We detect the signatures responsible for CIR/HSS-driven geomagnetic storms in all the available time series. 
    \item We associate the signatures belonging to the same event to each other. 
    \item We compare polarities from CHs on the Sun to the solar wind at L1 to evaluate the reliability of a polarity estimation. 
    \item We derive a function to approximate the solar wind speed from areas of CHs. \item We fit a periodic function to the ratios Dst/v and Kp/v separated by CH polarity on the day of the year (DOY).
    \item We combine the results to formulate a prediction model for Dst and Kp as a function of CH polarity, CH area and DOY. 
\end{enumerate}
In the following, we describe each of these steps in more detail.

\subsection{Event Detection} \label{Sect:EventDetection}
First, we build a dataset comprising the properties of CHs on the Sun, the associated HSSs, CIRs, and orientation of the IMF at L1, and the strength of the associated geomagnetic storms. To detect coronal holes, we look for peaks in the timeline of the areas that CHs cover within a meridional slice. We require that the CHs cover at least $\SI{5}{\percent}$, of the slice, and their coverage must be at least $\SI{3}{\percent}$, higher than the minimum coverage in the days around. We require a minimum time period between one peak and another of 3 days. This leads to 540 CHs area signatures. 
Following previous studies \citep{Owens2005,Owens2008,Reiss2016}, we identify peaks in the solar wind profiles, by requiring that the solar wind velocity peaks exceed $\SI{400}{km/s}$. We require their prominence, i.e. the vertical distance between a peak and its surrounding minimum, to be at least $\SI{60}{km/s}$. We require the minimum time period between two neighboring peaks to be 3 days. We reject peaks which are caused by interplanetary coronal mass ejections (ICMEs) employing the \citet{RichardsonCane2003} ICME list\footnote{\url{http://www.srl.caltech.edu/ACE/ASC/DATA/level3/icmetable2.htm}}. We further exclude all HSS peaks from our dataset which could be affected by close-by ICMEs, i.e., we remove all HSSs peaks in the 2 days preceding and following the ICME window.
This procedure results in 367 HSSs signatures. 
For the IMF signatures, we search for the local maximum between [-3, -0.5] days from the solar wind velocity peak. We take this time window because the IMF peaks in the CIR compression region, and the compression region preceeds the HSS region having the maximum velocity.
For Dst and Kp, we look for peaks between the IMF peak and 2 days following the matching solar wind velocity peaks. Dst and Kp can culminate within the CIR region, but as they are cumulative geomagnetic effects, we can find their peak also a few days later. 
When the local maxima or minima of either the IMF, Dst or Kp do not correspond to a peak value, we remove them from the dataset.
This results in 365 HSSs, IMF, Dst and Kp signatures. 

\begin{figure*}
\centering
	\includegraphics[width=17cm]{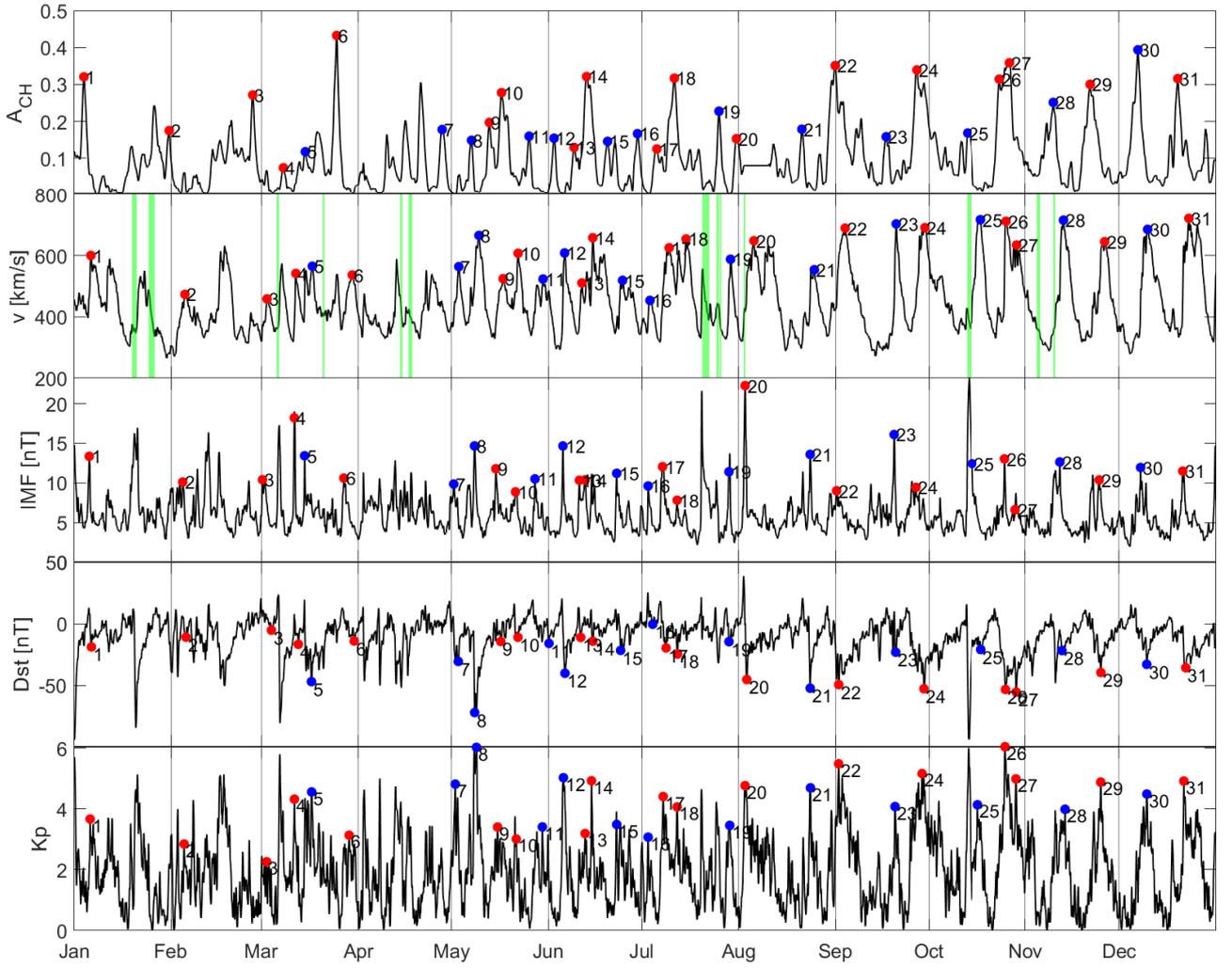}
    \caption{Event association for the year 2016. From top to bottom: CH area, A$_\textrm{CH}$, solar wind velocity, v, IMF, Dst, and Kp. Each marker followed by a number, shows the events that have completed the association scheme. Markers with the same number indicate the events that originate from the same peak in CH area, and produced a geomagnetic storm. The green bars in the solar wind speed subplot represent periods of ICME activity. Red markers show polarity away from the Sun, while the blue ones show polarity toward the Sun.}\label{Fig:panelplot2016}
\end{figure*}

Figure \ref{Fig:panelplot2016} shows the timelines of the CH areas, solar wind velocity, IMF, Dst, and Kp. The marked peaks are those just collected, and filtered by the next two steps to set up the forecasting model. 

\subsection{Event Association} \label{Sect:Event_Association}
Having the individual peaks in the CH areas, solar wind velocity, IMF, Dst and Kp, we need to relate them to causal chain of events. The event association requires that each peak of the solar wind velocity is associated with no more than one detected peak from the other parameters (CH area, IMF, Dst, Kp).
The signatures of the solar wind velocity and those of IMF, Dst and Kp have already been associated with each other during the detection process. To associate the signatures of the solar wind velocity to those of CHs area we make use of the following approach. We focus separately on each of the detected solar wind velocity signatures. We search for the closest peak in CH area within a time window of [-6, -2] days, measured from the identified solar wind velocity peak. At the end of this step, unpaired signatures are removed. Amongst the 365 peaks of the solar wind velocity and the 540 peaks of CH area, we associated 259 pairs. The number of the events decreases because not every CIR from CHs actually hits the Earth. 
For three CH events, the CH event have been associated to two different close-by solar wind events. We kept the events having the highest prominence in the solar wind events, reducing our data set to 256 pairs.
In total, we found 256 geomagnetic storm events having one-to-one correspondence in the signatures of the solar wind velocity, CH area, IMF, Dst, and Kp. 

\subsection{Comparison of the HSS polarity at L1 with the polarity of coronal holes}\label{Polarities_estimation}
Since the polarity of the IMF at L1 is related to the polarity of CHs within the central meridional slice, we want to check whether:
\begin{equation}
 \sign(P_\textrm{CH}) = \sign(P_\textrm{L1}).
\end{equation}
P$_\textrm{CH}$ is the polarity of the CHs, i.e., the ratio of the signed to unsigned vertical magnetic field strength derived from all CHs in the central meridian slice.
P$_\textrm{L1}$ is the magnetic polarity of the IMF. We define it following \cite{Neugebauer2002}, as the normalized IMF vector projected to the expected direction of the Parker spiral:
\begin{equation}
  P_{\textrm{L1}} = (B_r - \Omega R B_t/v\cos\lambda)/ [\sqrt{1+(\Omega R cos\lambda/v)^2} \sqrt{B_r^2 + B_t^2} ].  \label{eq:Neugebauer2002} 
\end{equation}
Here $B_{r}$ and $B_{t}$ are the radial and tangential component of the IMF, $\lambda$ is the latitudinal position of the spacecraft in HCI coordinates, v is the solar wind bulk velocity, $R = \SI{1}{AU}$ is the distance between Earth and Sun, and $\Omega = \SI{2.7e6}{\per\second}$ is the solar rotation rate.

According to our definitions, the polarities of the CHs and of the solar wind are both in the interval from $[-1,+1]$, where the sign refers to the predominant direction of the magnetic flux. Positive values correspond to a flux directed away from the Sun, and negative values toward the Sun.
As values close to zero do not show a predominant direction of the magnetic field, we classify values between $[-0.1,+0.1]$ as \textit{not defined}.
\begin{figure*}
\centering
	\includegraphics[width=17cm]{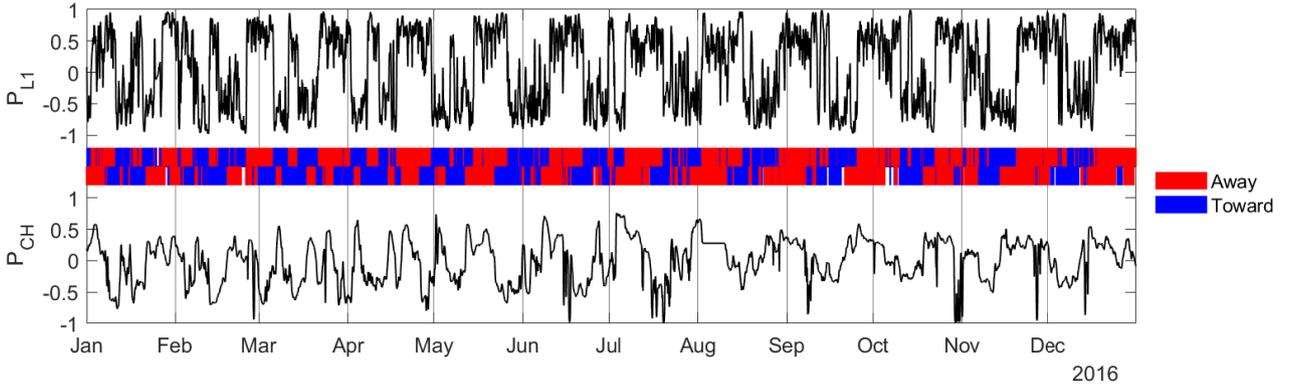}
	\caption{Polarity estimations for the year 2016.
	The top panel shows the polarity, P$_{\textrm{L1}}$, of the IMF derived at L1 using Equation~(\ref{eq:Neugebauer2002}). The bottom panel depicts the mean magnetic flux density of CH, P$_\textrm{CH}$. Red bars in the middle panel represent positive polarity (away from the Sun), while blue bars give negative polarity (toward the Sun), and white gaps indicate periods of \textit{not defined} polarity. The bar-plot on top refers to P$_{\textrm{L1}}$ and the one on the bottom to P$_\textrm{CH}$.}\label{Fig:polarity2016}
\end{figure*}
Figure~\ref{Fig:polarity2016} shows the derived polarity, P$_{\textrm{L1}}$, of the IMF at L1 (top panel) and the polarity of CHs within the central meridional slice, P$_\textrm{CH}$ (bottom panel) for the year 2016. The bar plots in the middle panel display the predominant polarity direction of P$_{\textrm{L1}}$ (top) and P$_\textrm{CH}$ (bottom). From the bar plots, we can see that the polarity on the Sun usually matches well the polarity at L1, preceding it by 2--6 days. This is consistent with the solar wind's propagation time from the Sun to L1.

Comparing the polarity of CH events on the Sun with the polarity of the IMF events at L1, we find that the polarity is conserved in around $\SI{83}{\percent}$ of events. This opens the possibility to use the magnetic field derived directly from solar observations instead of that at L1. Of the remaining $\SI{17}{\percent}$, less than $\SI{1}{\percent}$ comes from events with \textit{non defined} polarity. A high percentage of polarities does not match mostly because multiple coronal holes with different polarities are in the central meridional slice at the same time. 
All the events where the polarity between L1 and that on the Sun do not match or it is \textit{not defined} are removed from the further analysis. This reduces our dataset to 212 events.

\subsection{Gaussian process regression for predicting velocity peaks from CH area}
To account for the effect of the solar wind velocity on the Dst and Kp indices, we first fit an empirical relationship that links the solar wind velocity to the CHs area on the Sun. We fit the GPR model with the constraint of having a velocity of $\SI{400}{km/s}$ at zero CHs area, which corresponds to the velocity of the slow solar wind. We employ as prior covariance function of the GPR the radial basis function (RBF, Equation~(\ref{eq:rbf_kernel})).

The predictive Gaussian process (GP) of the solar wind velocity has the form:
     \begin{equation}
     v(A_\textrm{CH}) \sim GP(\,m_v(A_\textrm{CH}),\, \sigma_v(A_\textrm{CH})),
     \label{eq:V(A)}
    \end{equation}
and it depends only on the values of CH area.
Here, $m_{v}$ and $\sigma_{v}$ are the predicted mean function and predicted standard deviation function of $v(A_\textrm{CH})$. The fitted hyperparameters are shown in Table~\ref{tab:hyperparameters}. 

\begin{figure}
   \centering
    {\includegraphics[width=1\columnwidth]{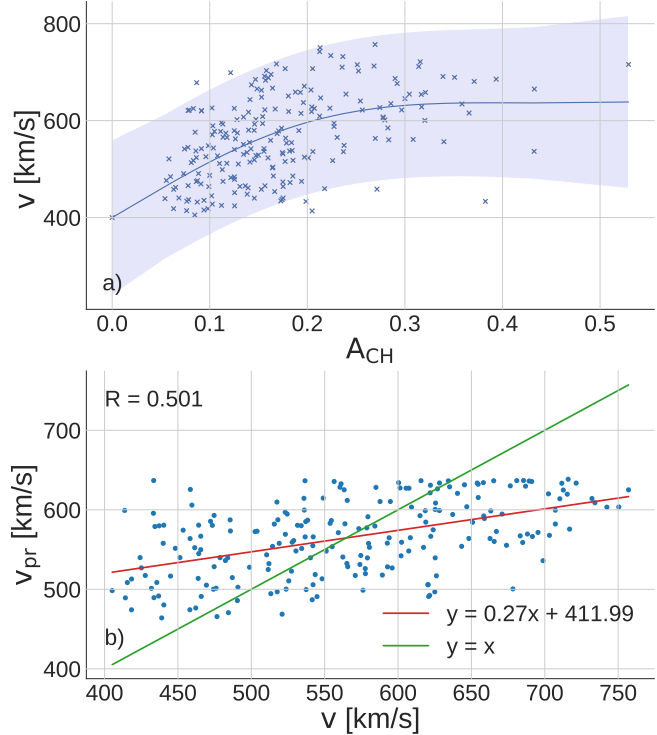}}
    \caption{ Panel (a) shows the solar wind velocity, v, versus CH area, $A_\textrm{CH}$. The markers are the events extracted in our study period, used for the GPR fitting. The solid line is the mean function of the predictive GP (Equation~\ref{eq:V(A)}), and the shaded region depicts the interval where it is possible to find a realization of the solar wind velocity from the CH area with $\SI{95}{\percent}$ confidence. Panel (b) is the scatter plot of predicted solar wind velocity peaks (v$_\textrm{pr}$) against observed solar wind velocity events (v) in km/s. Red and green lines represent the linear least-squares fit, and the y = x line, respectively. R is the correlation coefficient among predicted and observed peaks.
    }\label{fig:HSS(A) corr}
\end{figure}

Figure~\ref{fig:HSS(A) corr}(a) shows the solar wind peak velocity versus the CH area. The solid line shows the predicted mean function, $m_v$, from the GP fitting, and the shaded region depicts the GP's uncertainty, corresponding to the $\SI{95}{\percent}$ confidence interval.
The fitted function shows a rise of $m_v$ from $400$ to approximately $\SI{600}{km/s}$ corresponding to an increase of CH area from $0$ to $0.3$. After that we see the saturation of $m_v$ around $\SI{600}{km/s}$.
Figure~\ref{fig:HSS(A) corr}(b) shows the predicted versus the measured solar wind velocity peaks. On the $x$-axis there are the observed solar wind events, v, and on the $y$-axis the solar wind peaks predicted, v$_\textrm{pr}$. The red line is a linear least-square fit to the data and the green line is the perfect one-to-one correspondence with v $=$ v$_\textrm{pr}$. We see that the slope of the linear fit is much lower than $1$, i.e. the slope of the one-to-one correspondence. This means that the fitted model produces a lower rate of change in v$_\textrm{pr}$ with increasing A$_\textrm{CH}$ than the one in the actual measurements. The forecast verification metrics used to assess the results are the Pearson correlation coefficient R, the mean error ME, the absolute mean error MAE, and the root mean squared RMSE computed between v$_\textrm{pr}$ and v, are given in Table~\ref{tab:ForecastVer_v}. We see the Pearson coefficient is R = 0.50, showing a moderate correlation between v$_\textrm{pr}$ and v.
The MAE and RMSE are 62 km/s and 76 km/s, respectively, while the velocities of the HSSs in our dataset lie in a range of 200 km/s, i.e., from 450 km/s to 650 km/s. As a result of the strong scatter in the solar wind velocity predictions, we also expect a large scatter in the following Dst and Kp predictions.
\begin{table}[]
	\centering
	\caption{ \label{tab:hyperparameters} Optimal hyperparameters of the the GPR model for the solar wind velocity (top) and the ratios Dst/v and Kp/v separated by the CH polarity (bottom).}
     \begin{tabular}{c l c c c} 
      \hline
   	  	  &  &$\sigma_K$ & $l_K$ & $\sigma_\textrm{n}$ \\ 
   	  	 \midrule
        \multirow{2}{*}{v} &
           &
          \multirow{2}{*}{2.25} &
          \multirow{2}{*}{0.27} &
          \multirow{2}{*}{0.87} \\
          \\
     	 \midrule
     	 \multirow{2}{*}{Dst/v} 
      	    & Toward &0.49 & 0.44 & 0.88 \\
     	    & Away  &0.70 & 0.44 & 0.76  \\
     	 \midrule
     	 \multirow{2}{*}{Kp/v}  
     	      & Toward &0.38 & 0.45 & 0.93 \\
     	      & Away  &0.56 & 0.28 & 0.82 \\
     	 \midrule
 	 \end{tabular}
\end{table}

\begin{table}[]
\centering
\caption{Results of the forecast verification of the the GPR model for the solar wind velocity (top) and the ratios Dst/v and Kp/v separated by the CH polarity (bottom).}
\label{tab:ForecastVer_v}
\resizebox{\textwidth}{!}{%
\begin{tabular}{@{}ccccccc@{}}
\toprule
                     &        & R    & Slope & ME           & MAE         & RMSE        \\ \midrule
\multirow{2}{*}{v} & \multirow{2}{*}{} & \multirow{2}{*}{0.50} & \multirow{2}{*}{0.27} & \multirow{2}{*}{-1.47} & \multirow{2}{*}{62.24} & \multirow{2}{*}{75.74} \\     
\\ \midrule

\multirow{2}{*}{Dst/v} & Toward & 0.52 & 0.23  & $\SI{-7e-5}{}$ & 0.02        & 0.02        \\
                     & Away   & 0.68 & 0.42  & $\SI{-8e-5}{}$ & 0.01        & 0.02        \\     	 \midrule
\multirow{2}{*}{Kp/v}  & Toward & 0.42 & 0.14  & $\SI{6e-6}{}$  & $\SI{1e-3}{}$ & $\SI{1e-3}{}$ \\
                     & Away   & 0.61 & 0.32  & $\SI{7e-6}{}$  & $\SI{8e-4}{}$ & $\SI{1e-3}{}$ \\     
 \bottomrule
\end{tabular}%
}
\end{table}

\begin{table}[]
\centering
\caption{Results of the forecast verification computed using v$_\textrm{pr}$ and v between observed geomagnetic indices and their predictions separated by the CHs polarity.}
\label{tab:ForecastVer_dst_kp}
\resizebox{\textwidth}{!}{%
\begin{tabular}{@{}clllllll@{}}
\toprule
                          &                      &        & R    & Slope & ME    & MAE  & RMSE  \\ \midrule
\multirow{4}{*}{v$_\textrm{pr}$} & \multirow{2}{*}{Dst} & Toward & 0.63 & 0.26  & 0.43  & 9.64 & 12.39 \\
                          &                      & Away   & 0.73 & 0.43  & 0.73  & 8.70 & 12.18 \\
                          & \multirow{2}{*}{Kp}  & Toward & 0.65 & 0.26  & 0.02  & 0.54 & 0.77  \\
                          &                      & Away   & 0.67 & 0.39  & -0.04 & 0.57 & 0.71  \\ \midrule
\multirow{4}{*}{v}         & \multirow{2}{*}{Dst} & Toward & 0.69 & 0.36  & 0.39  & 8.77 & 11.43 \\
                          &                      & Away   & 0.76 & 0.52  & 0.06  & 8.16 & 11.45 \\
                          & \multirow{2}{*}{Kp}  & Toward & 0.71 & 0.52  & 0.01  & 0.54 & 0.67  \\
                          &                      & Away   & 0.80 & 0.64  & 0.01  & 0.46 & 0.57  \\ \bottomrule
\end{tabular}%
}
\end{table}

\subsection{Gaussian process regression for predicting Dst/v and Kp/v}
The angle between Earth’s magnetic field axis and the IMF along the Archimedian spiral varies over the year. This affects Bs and, in turn, Dst/v and Kp/v. Since Bs separated by polarity shows a \textit{pair of spectacles} pattern, we fit the GPR model on the events Dst/v and Kp/v separated by polarity employing a sinusoidal function. We, hence, employ as prior a periodic kernel with a fixed periodicity of 365.24 days (Equation~\ref{eq:periodic_kernel}).
This results in four functions, separately for the events with polarity away from the Sun, $\textrm{Dst}_+/\textrm{v}_+$ and $\textrm{Kp}_+/\textrm{v}_+$, and for events with polarity towards the Sun, $\textrm{Dst}_-/\textrm{v}_-$ and $\textrm{Kp}_-/\textrm{v}_-$. 
Considering as example the ratio $\textrm{Dst}_+/\textrm{v}_+$, its predictive GP is:
     \begin{equation}
     \begin{split}
     \frac{Dst_+}{v_+}(DOY) \sim GP(\,m_{D+}(DOY),\, \sigma_{D+}(DOY)),
     \label{g(DOY)}
     \end{split}
    \end{equation}
where $m_{D+}$ and $\sigma_{D+}$ are the functions of the mean and the standard deviation dependent on the DOY. 
The optimal hyperparameters for each of the four models are presented in Table~\ref{tab:hyperparameters}.
\begin{figure*}
        \centering
        \begin{subfigure}[b]{0.49\textwidth}
            \includegraphics[width=\linewidth]{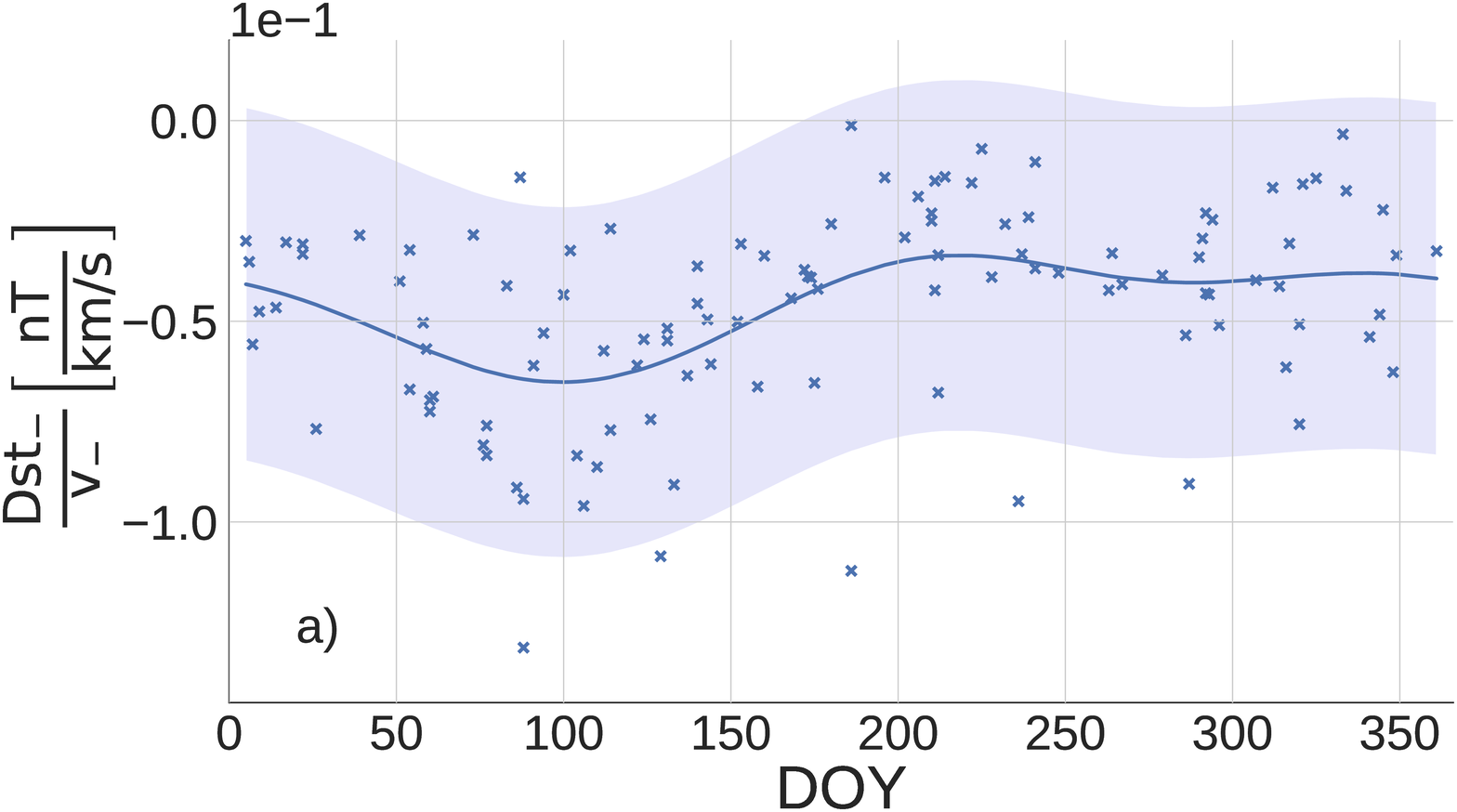}  
        \end{subfigure}
        \begin{subfigure}[b]{0.49\textwidth}
            \includegraphics[width=\linewidth]{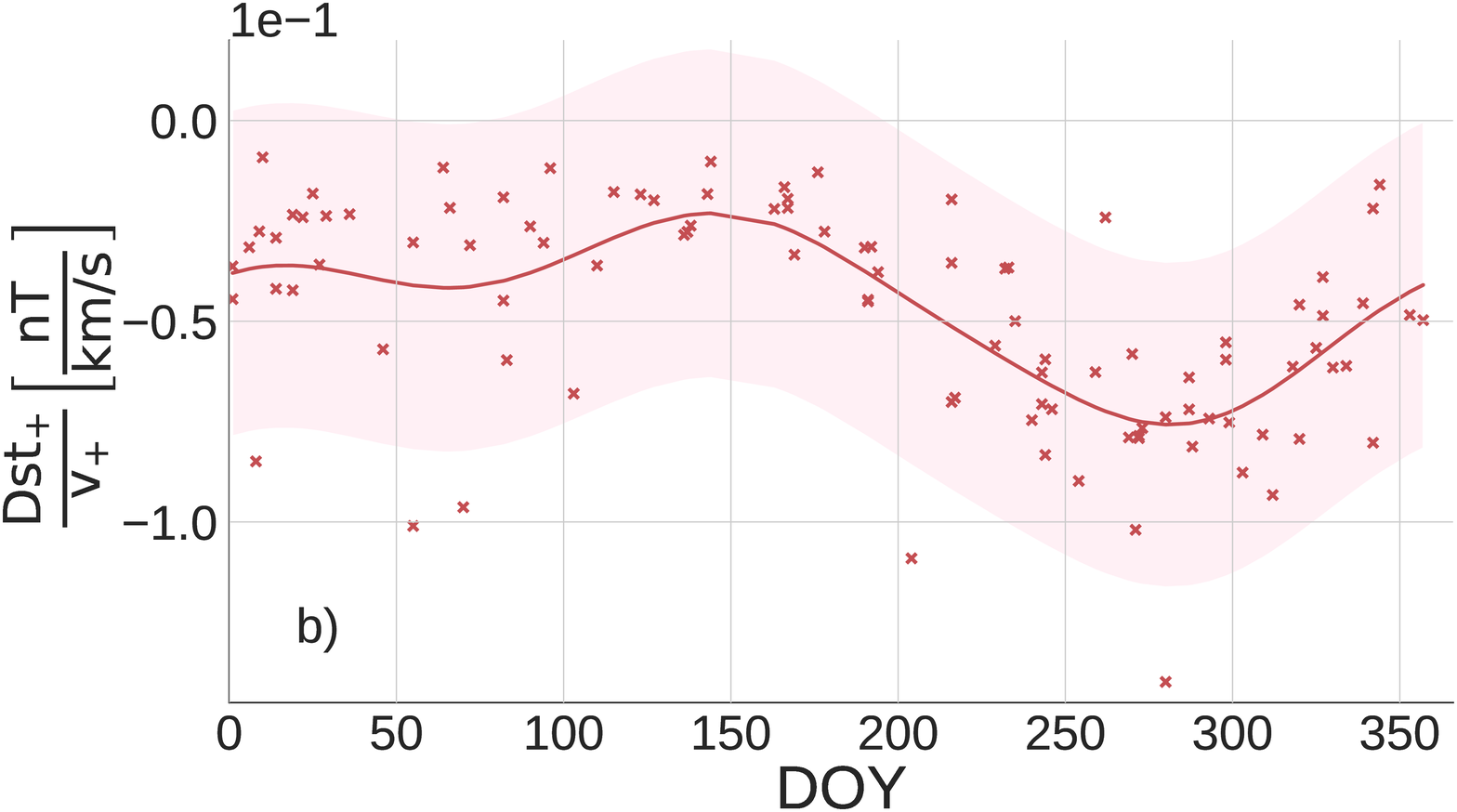}  
        \end{subfigure}
        \centering
        \begin{subfigure}[b]{0.49\textwidth}
            \includegraphics[width=\linewidth]{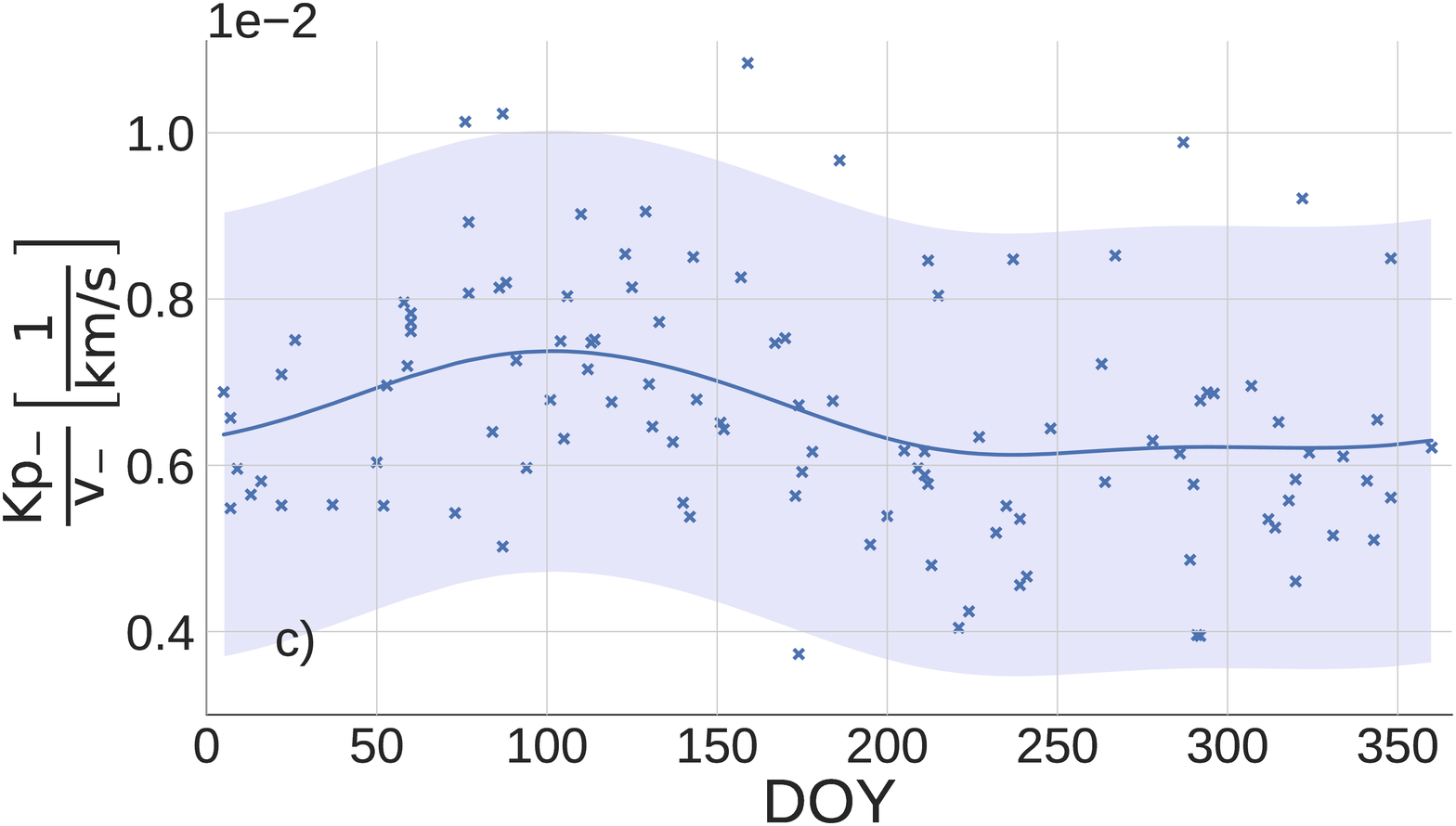}  
        \end{subfigure}
        \begin{subfigure}[b]{0.49\textwidth}
            \includegraphics[width=\linewidth]{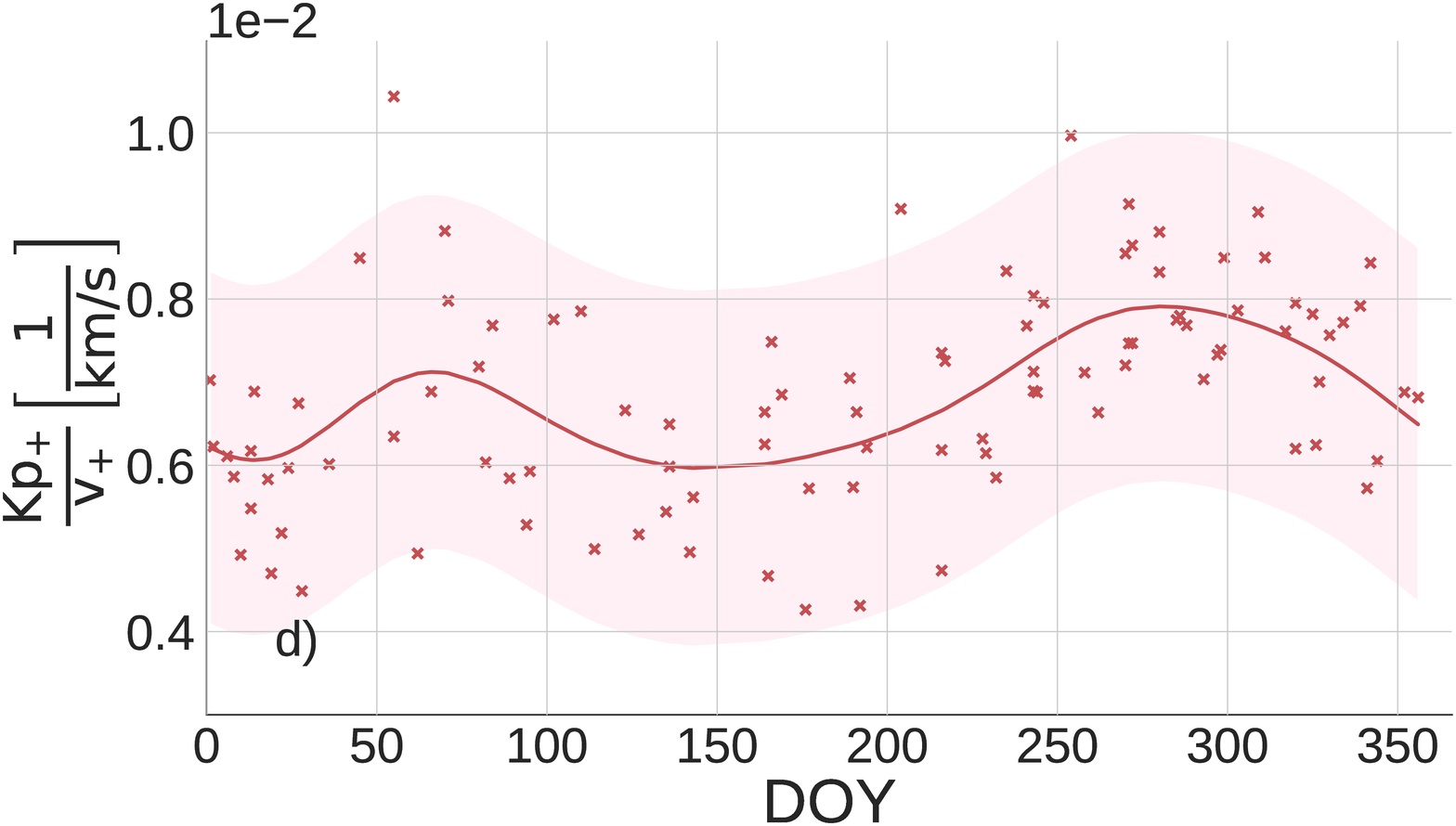}  
        \end{subfigure}
        \caption{
        Ratio of Dst and Kp over the solar wind velocity, v, versus DOY when the events in the considered geomagnetic index took place. The markers are the observations extracted in our study period. The solid line is the mean function of the predictive GP, whereas the shaded region depicts the interval where it is possible to find a realization of the GP with $\SI{95}{\percent}$ of confidence. 
        Panels (a) and (b) show the predictive GP of Dst fitted on the events with CH polarity toward and away from the Sun, respectively.  
        Panels (c) and (d) show the predictive GP of Kp fitted on the events with toward and away polarity, respectively.
        }
        \label{fig:GPR}
\end{figure*}    

Figure~\ref{fig:GPR}(a) and \ref{fig:GPR}(b) belong to Dst/v and Figure~\ref{fig:GPR}(c) and \ref{fig:GPR}(d) belong to Kp/v, and shows their relationship versus the DOY. The solid line shows the predicted mean function from the GP fitting, and the shaded region depicts the $\SI{95}{\percent}$ confidence interval of the GP. Figure~\ref{fig:GPR}(a), related to the CH polarity toward the Sun, and \ref{fig:GPR}(b), related to the CH polarity away from the Sun, show an almost mirror-like pattern for the measurement of Dst/v, with an annual period. Events with CH polarity toward the Sun have the strongest Dst drops normalized to the solar wind velocity around the spring equinox, while the Dst/v ratio is rather constant for the rest of the year. CH events with polarity away from the Sun show the same pattern, but with the strongest Dst/v drops around the fall equinox.
Figures~\ref{fig:GPR}(c)~and~\ref{fig:GPR}(d) show a similar trend, whereby maxima in Kp/v are related to minima in Dst/v. This is because the Kp index shows the most severe activity at its peaks, while the Dst index shows the strongest geomagnetic activity at its drops. 

In Table \ref{tab:ForecastVer_v}, we observe that Dst/v and Kp/v with polarities away from the Sun are better predicted than with polarities toward the Sun. In particular, Dst/v has a correlation coefficient of 0.68 when the polarity is away and 0.52 when it is toward the Sun, whereas the MAE is 0.01~nT/(km/s) when the polarity is away and 0.02~nT/(km/s) when it is toward the Sun. For Kp/v, the difference in the correlation coefficient R is 0.19, being similar to the Dst/v case. The MAE, on the other hand, differs by \num{2d-4}~1/(km/s), being \num{1d-3}~1/(km/s) for polarities toward the Sun and \num{8d-4}~1/(km/s) for polarities away from the Sun.

For the following analysis, we denote the predictions of $\textrm{Dst}_+/\textrm{v}_+$ as $(Dst_+/v_+)_{pr}$, and we use the same notation also for predictions of $\textrm{Dst}_-/\textrm{v}_-$, $\textrm{Kp}_+/\textrm{v}_+$ and $\textrm{Kp}_-/\textrm{v}_-$.

\subsection{Forecasting geomagnetic indices Dst and Kp}
To predict the Kp and Dst indices, we use the results of the previous section. As input we use the CH area measured directly at the Sun, the CHs polarity within the central meridional slice, and the DOY when Dst and Kp events take place. To predict the Dst and Kp indices, we multiply the predictions of $\textrm{Dst}_+/\textrm{v}_+$, $\textrm{Dst}_-/\textrm{v}_-$, $\textrm{Kp}_+/\textrm{v}_+$ and $\textrm{Kp}_-/\textrm{v}_-$ by the predicted peak values of solar wind velocity, v$_\textrm{pr}$, from the CH area:

\begin{equation}
\begin{aligned}[c]
(Dst_+)_{pr} = \left(\frac{Dst_+}{v_+}\right)_{pr} \cdot v_{pr};\\
(Kp_+)_{pr} = \left(\frac{Kp_+}{v_+}\right)_{pr} \cdot v_{pr};
\end{aligned}
\qquad
\begin{aligned}[c]
  (Dst_-)_{pr} = \left(\frac{Dst_-}{v_-}\right)_{pr} \cdot v_{pr};\\
  (Kp_-)_{pr} = \left(\frac{Kp_-}{v_-}\right)_{pr} \cdot v_{pr}.
\end{aligned}
\label{eq:forecasting model}
\end{equation}

\begin{figure*}
    \centering
    \begin{subfigure}[b]{0.49\textwidth}
        \includegraphics[width=\linewidth]{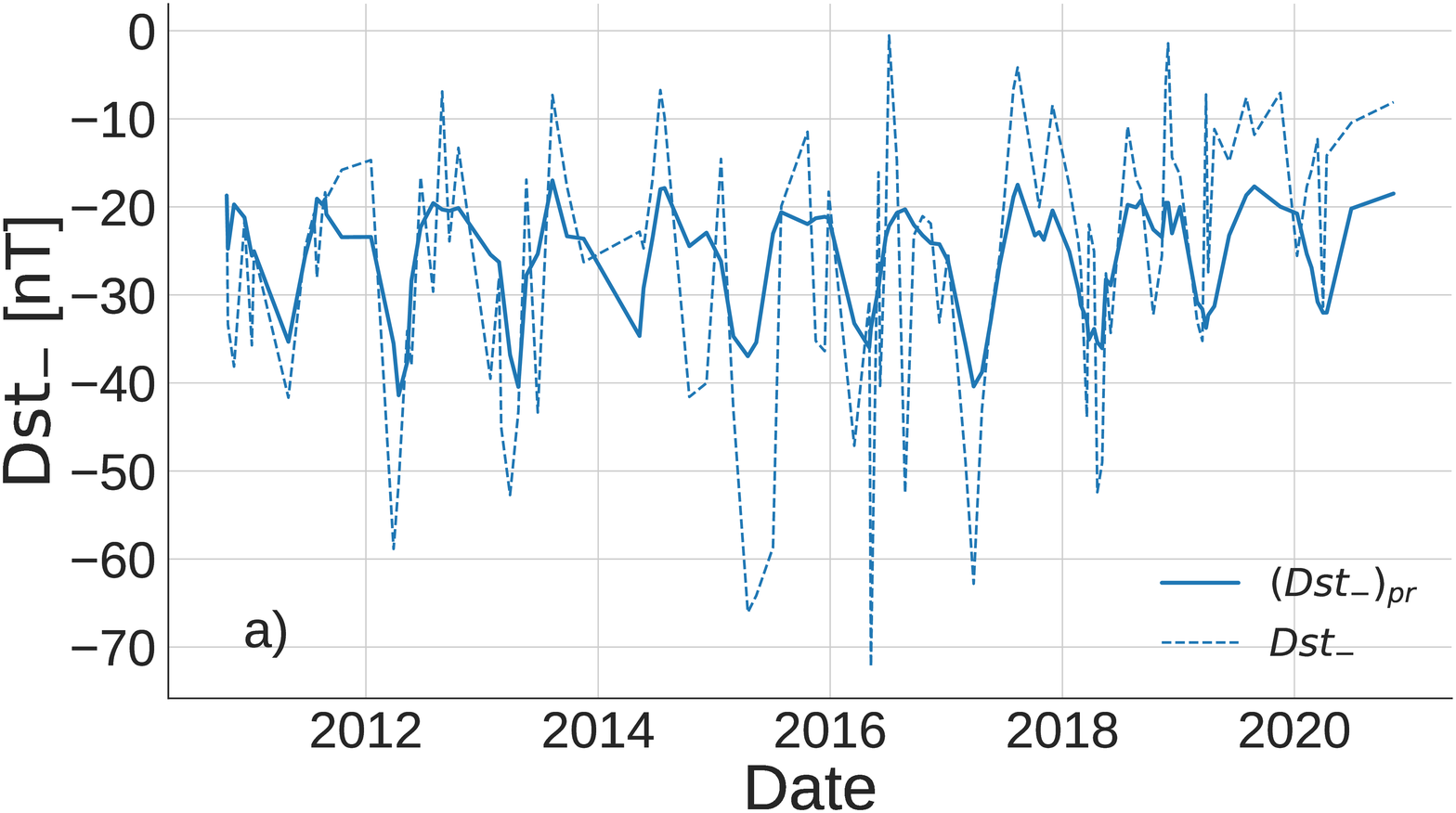}  
    \end{subfigure}
    \begin{subfigure}[b]{0.49\textwidth}
        \includegraphics[width=\linewidth]{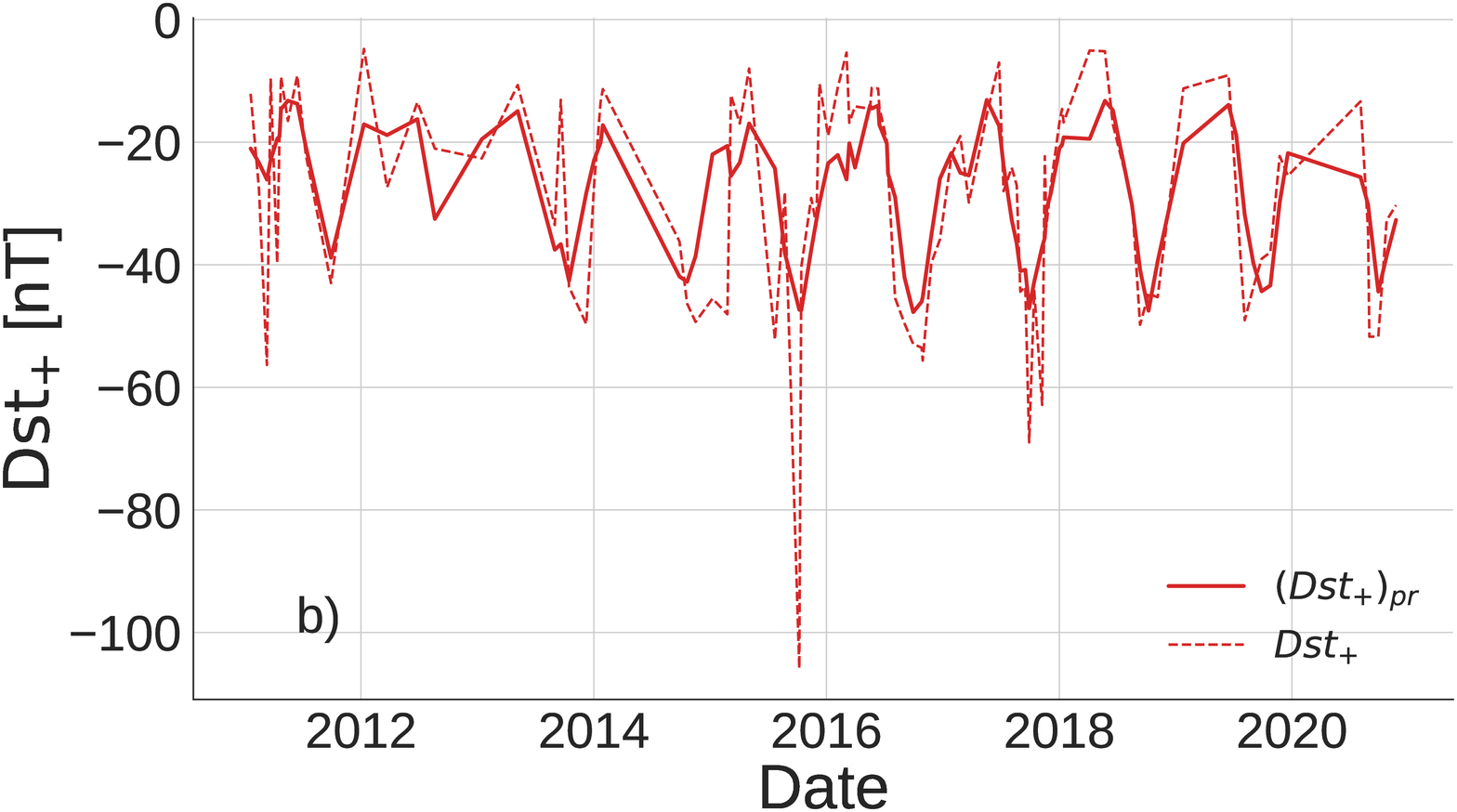}  
    \end{subfigure}
    
    \centering
    \begin{subfigure}[b]{0.49\textwidth}
        \includegraphics[width=\linewidth]{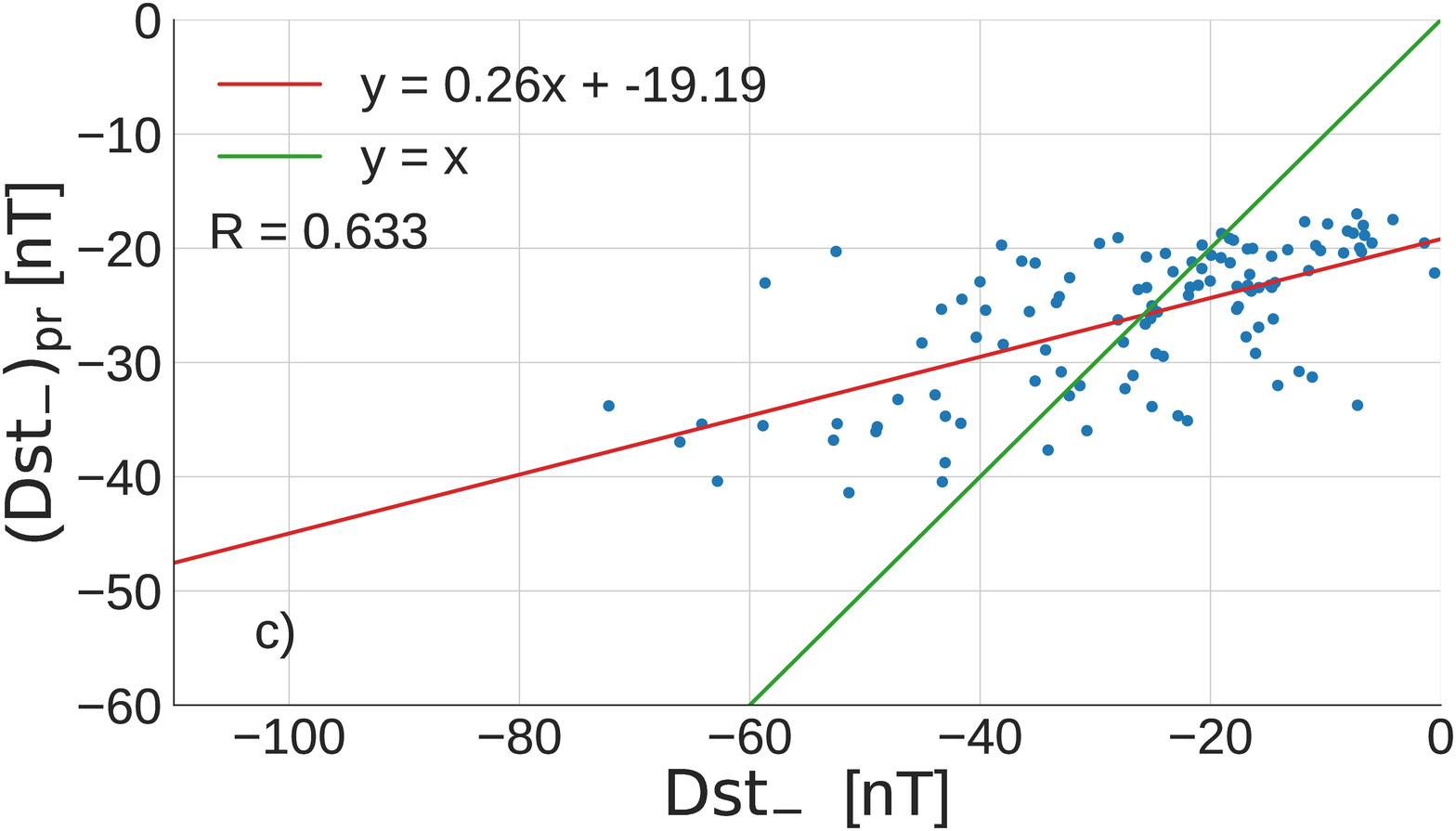}  
    \end{subfigure}
    \begin{subfigure}[b]{0.49\textwidth}
        \includegraphics[width=\linewidth]{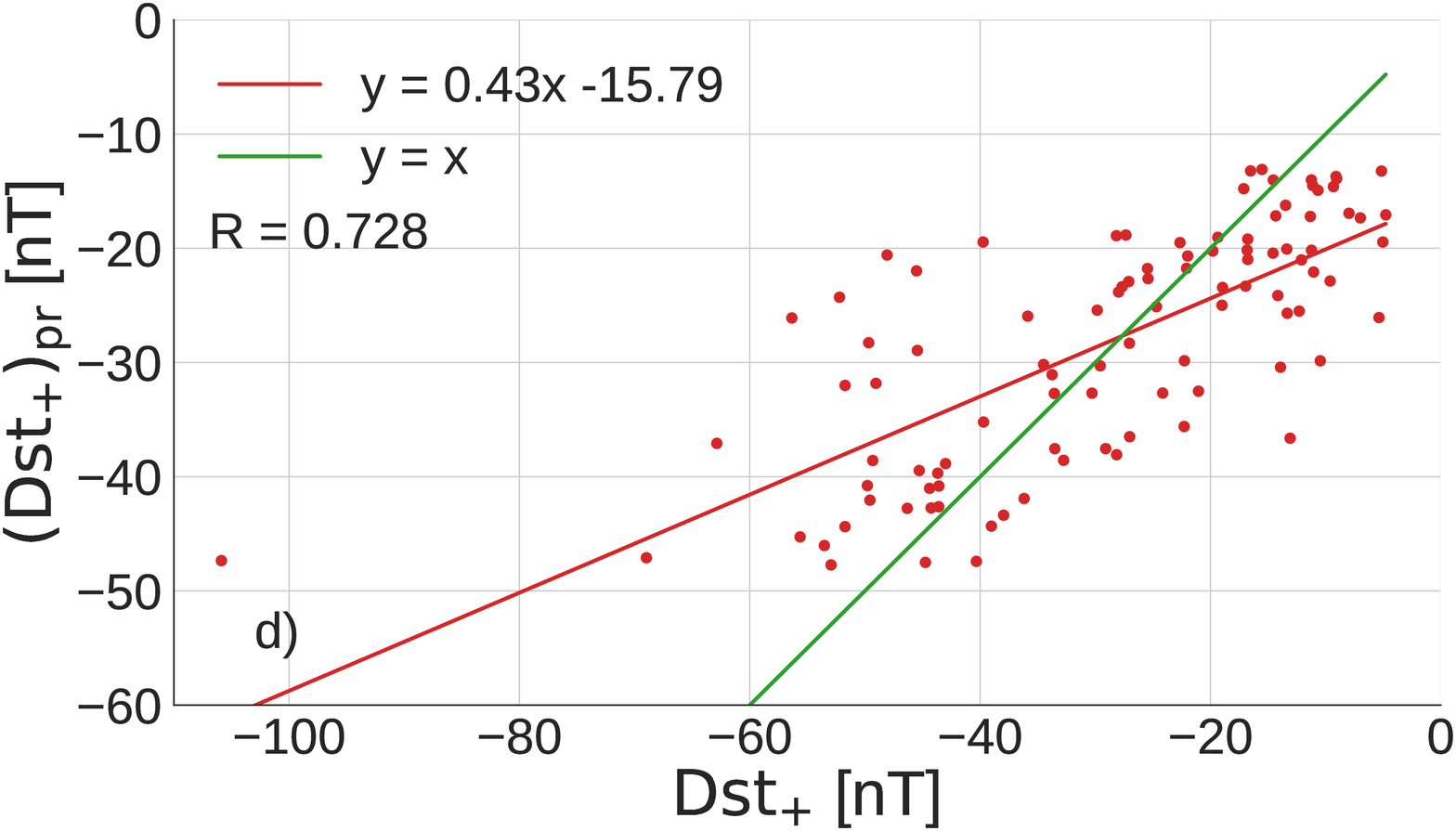}  
    \end{subfigure}
    \caption{
    Comparison between predicted and observed peaks of Dst.
    Panels (a) and (b) show observed (dashed line) and predicted (solid line) Dst peaks as function of time with CH polarity toward (blue) and away (red) from the Sun. Panels (c) and (d) give scatter plots between predicted  (y-axis) and observed Dst peaks
    ($x$-axis). Red and green lines represent the linear least-squares fit, described by the function in the top-left corner, and the $y = x$ line, respectively. R is the correlation coefficient between the predicted and observed quantities.} \label{fig:Dstpred}
\end{figure*} 
  
\begin{figure*}
    \centering
    \begin{subfigure}[b]{0.49\textwidth}
        \includegraphics[width=\linewidth]{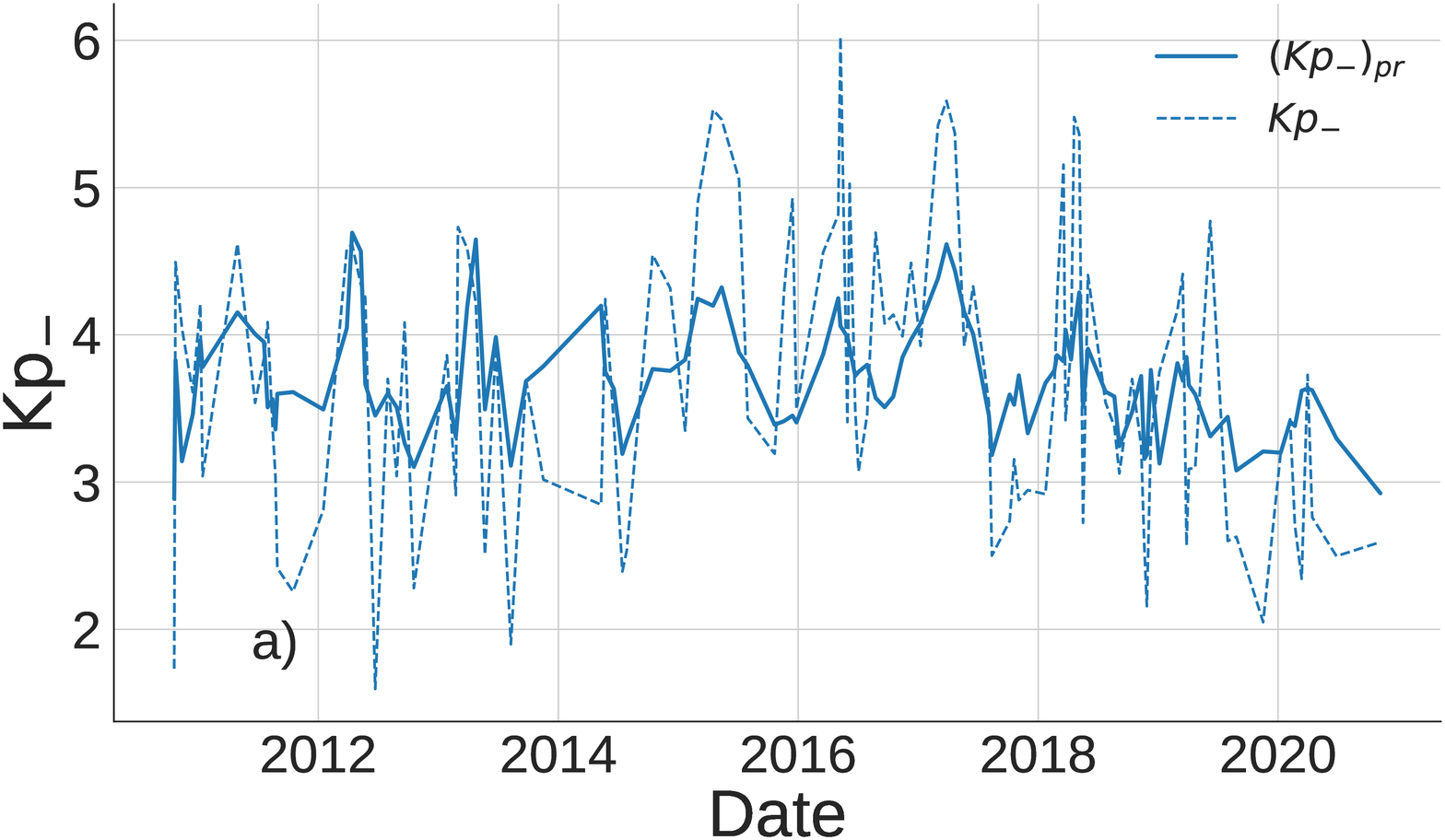}  
    \end{subfigure}
    \begin{subfigure}[b]{0.49\textwidth}
        \includegraphics[width=\linewidth]{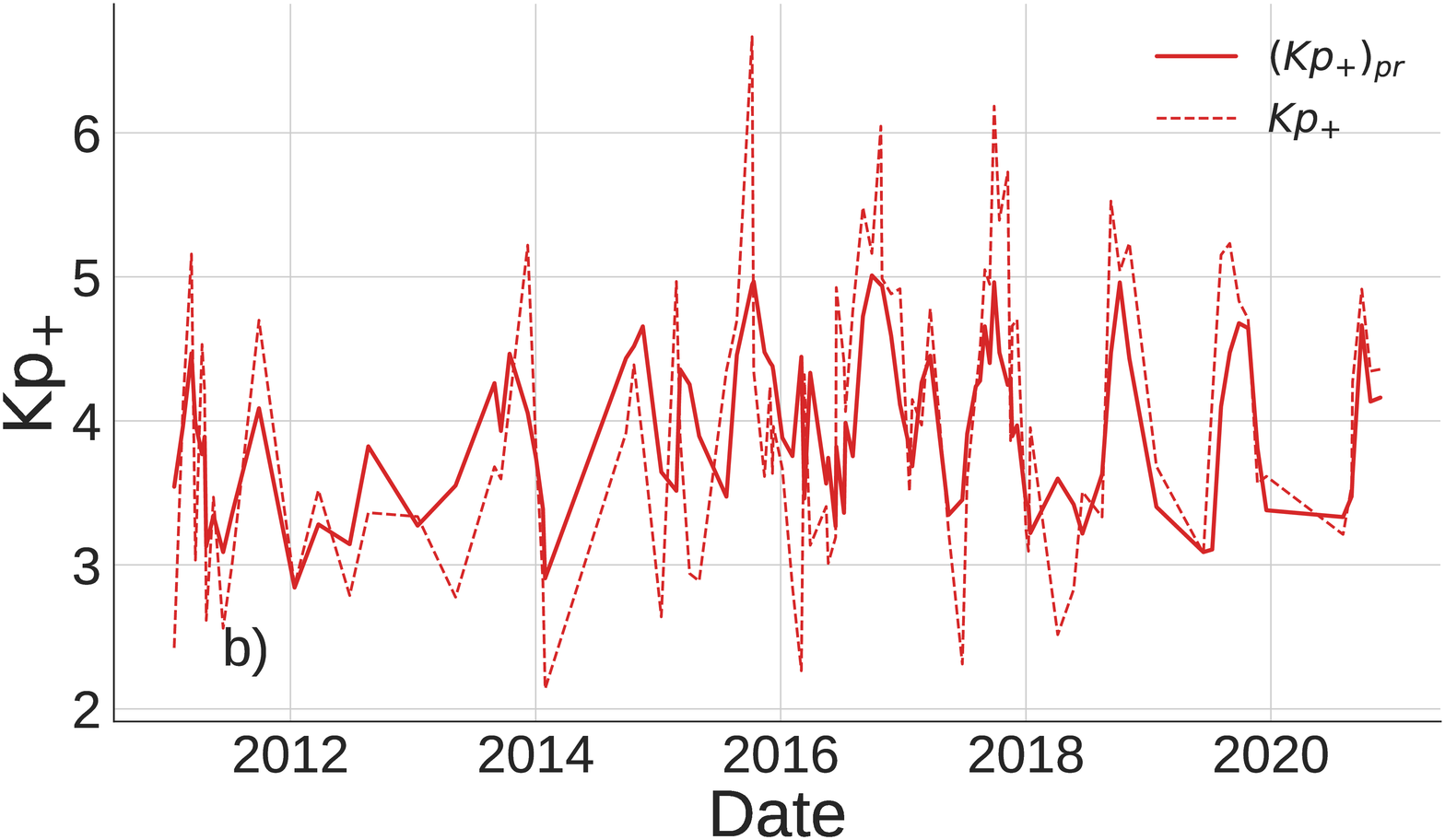}  
    \end{subfigure}
    
    \centering
    \begin{subfigure}[b]{0.49\textwidth}
        \includegraphics[width=\linewidth]{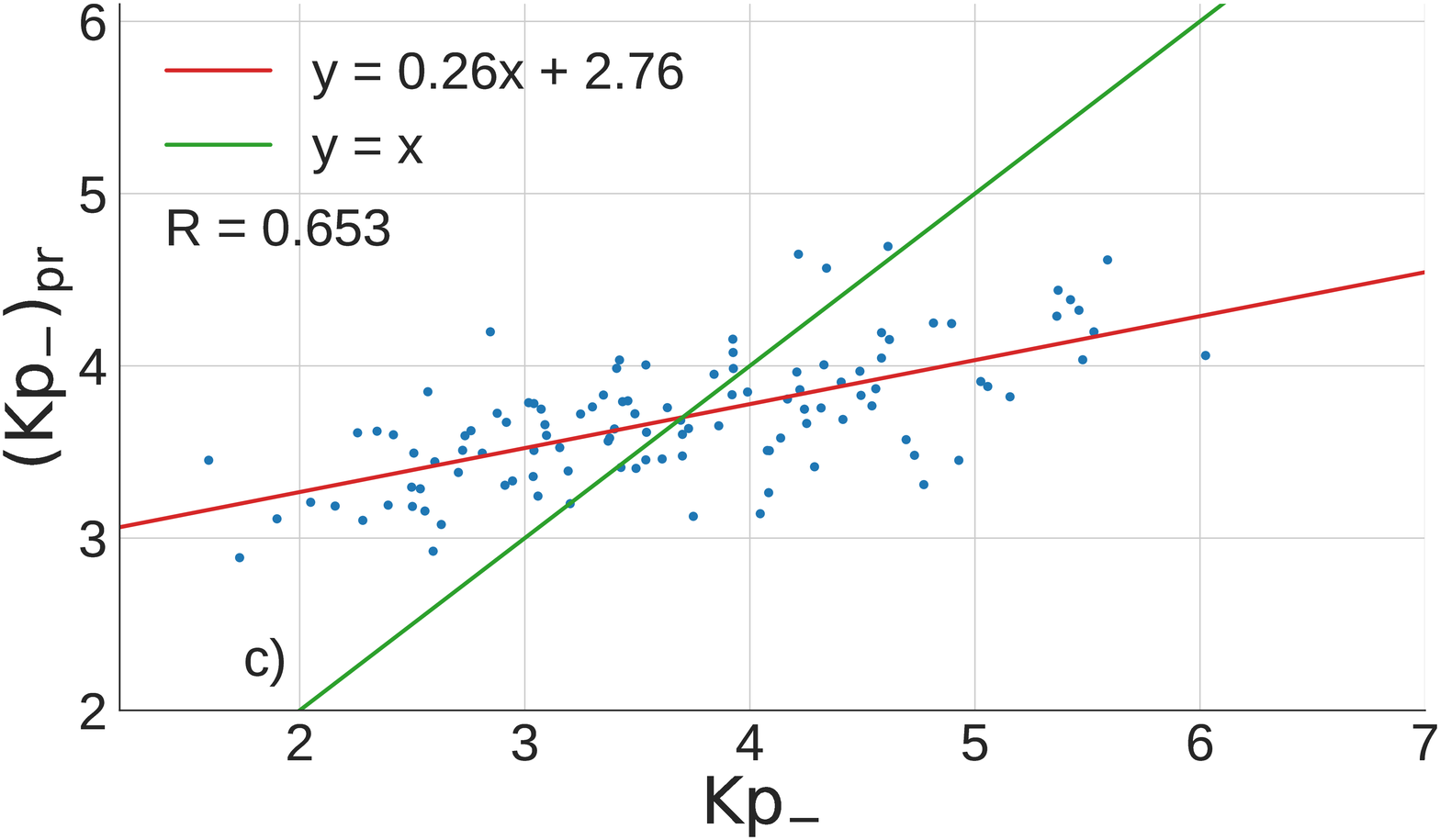}  
    \end{subfigure}
    \begin{subfigure}[b]{0.49\textwidth}
        \includegraphics[width=\linewidth]{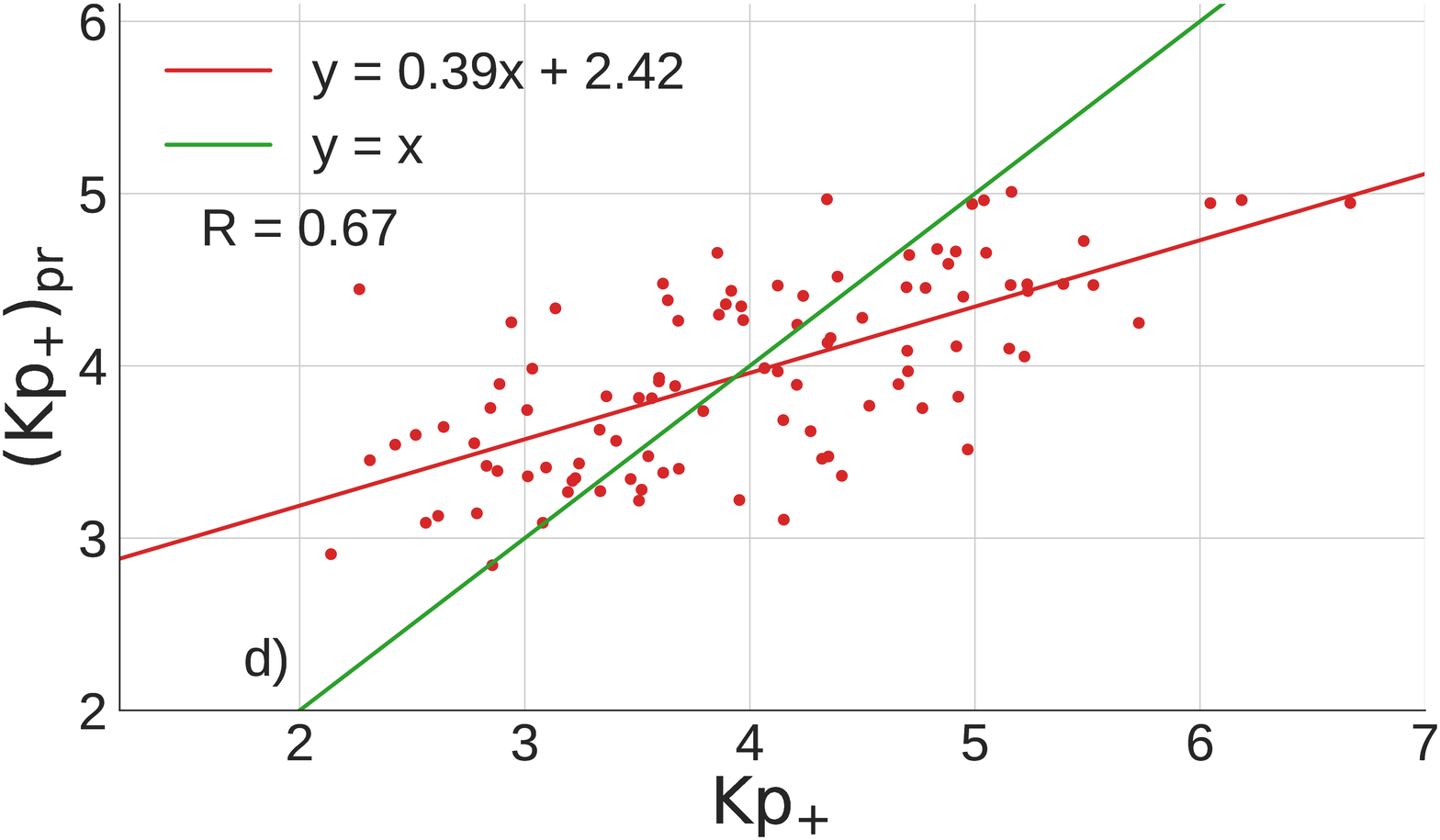}  
    \end{subfigure}
    \caption{
    The same as Figure~\ref{fig:Dstpred}, but for the Kp index.
    }
    \label{fig:Kppred}
\end{figure*} 

Figure~\ref{fig:Dstpred}(a)~and~\ref{fig:Dstpred}(b) show the predicted Dst$_\textrm{pr}$ (solid line) and observed Dst peaks (dashed line) as function of time. Thereby, Figure~\ref{fig:Dstpred}(a) shows the temporal evolution for peaks with CH polarity directed towards the Sun, and \ref{fig:Dstpred}(b) away from the Sun. 
Figures~\ref{fig:Dstpred}(c)~and~\ref{fig:Dstpred}(d) present the scatter plot between the predicted Dst$_\textrm{pr}$ and observed Dst peaks. Red and green lines indicate the linear least-squares fits, and the one-to-one correspondence, respectively. The plots are divided by polarity; data with polarity away from the Sun are in the right panel, while those with polarity toward the Sun are in the left panel.

In Figures \ref{fig:Dstpred}(a) and \ref{fig:Dstpred}(b), we see that the annual periodicity we impose over Dst/v separated by the CH polarity makes the prediction curve adapt to the oscillatory evolution of the Dst measurements throughout the study period. In Figure~\ref{fig:Dstpred}(a), the predicted Dst peaks follow the measured Dst peaks but are underestimated. In Figure~\ref{fig:Dstpred}(c), observed peaks vary approximately between $[-70,0]$~nT, whereas the predicted peaks are concentrated close to the average value of Dst, in the narrower range $[-40, -20]$~nT. The slope of the line describing the trend is flat, being equal to $0.26$, while the correlation coefficient is R $= 0.63$. In Figure \ref{fig:Dstpred}(b), the periodic pattern of Dst peaks is more regular than in Figure \ref{fig:Dstpred}(a), and the algorithm better approximates Dst$_\textrm{pr}$. This is reflected in the scatter plot in  Figure~\ref{fig:Dstpred}(d), where the slope is higher, being $0.43$, and the correlation coefficient is R $= 0.73$. 

Figure~\ref{fig:Kppred} shows the comparison between observed Kp and its predictions, Kp$_\textrm{pr}$, following the same scheme as Figure~\ref{fig:Dstpred}. In Figures~\ref{fig:Kppred}(a) and \ref{fig:Kppred}(b), values of Kp$_\textrm{pr}$ follow the temporal evolution of the observed Kp. In Figure~\ref{fig:Kppred}(c), the slope is $0.26$, i.e., flat, and the correlation coefficient is R $= 0.65$. 
In Figure~\ref{fig:Kppred}(d), the slope is slightly steeper, being  $0.39$, and the correlation coefficient is R $= 0.67$.

\begin{figure*}
    \centering
    \begin{subfigure}[b]{0.49\textwidth}
        \includegraphics[width=\linewidth]{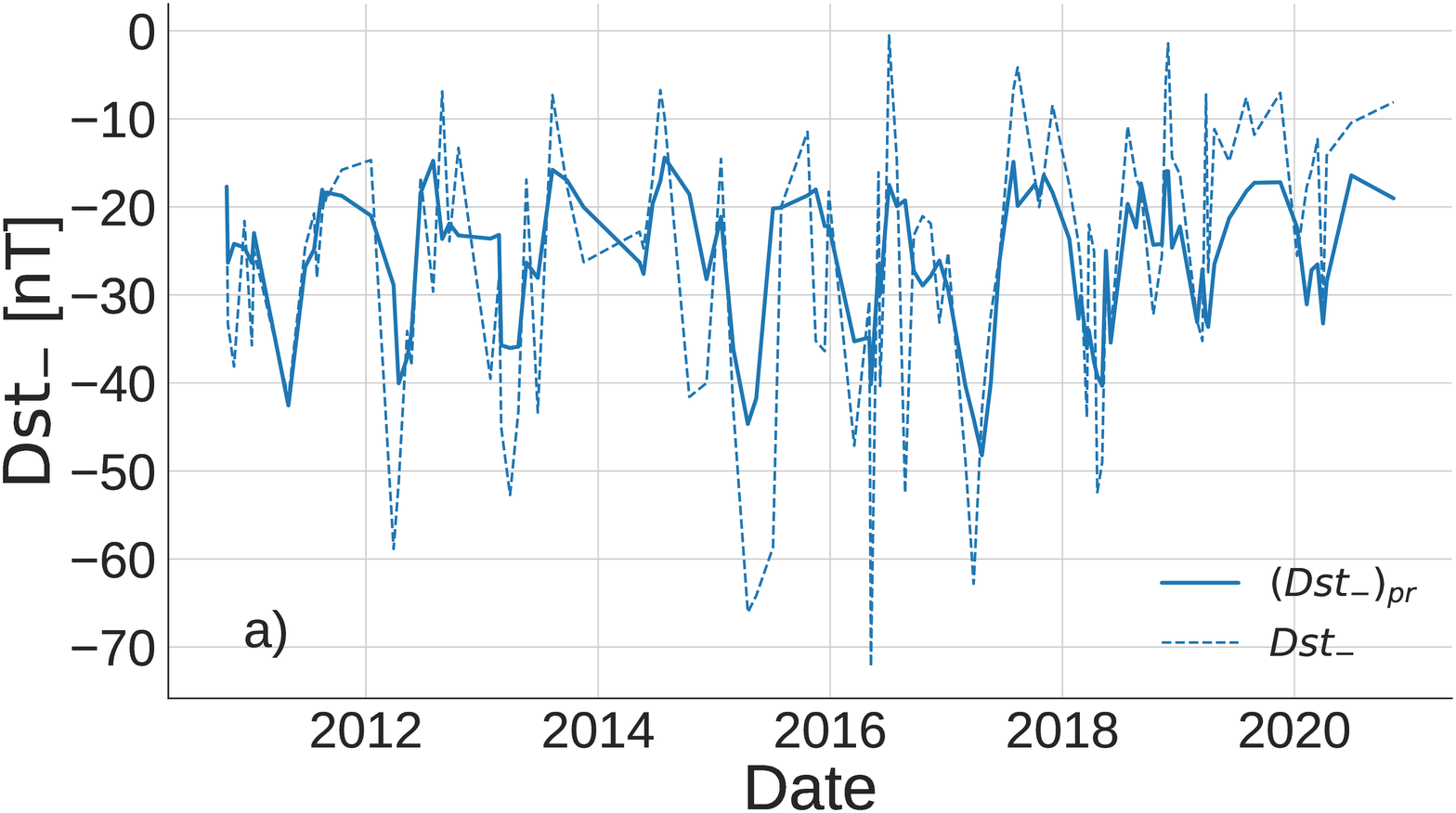}  
    \end{subfigure}
    \begin{subfigure}[b]{0.49\textwidth}
        \includegraphics[width=\linewidth]{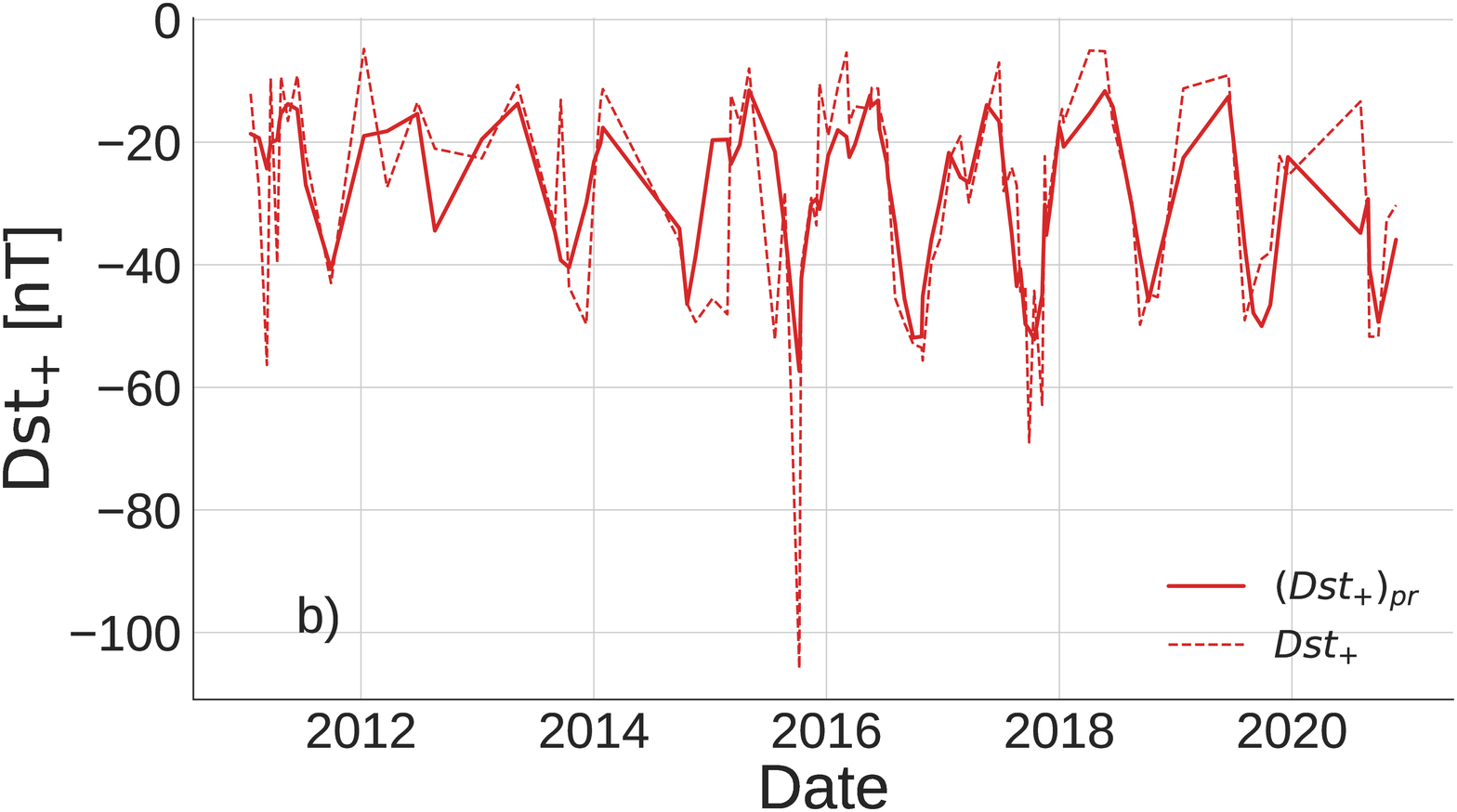}  
    \end{subfigure}
    
    \centering
    \begin{subfigure}[b]{0.49\textwidth}
        \includegraphics[width=\linewidth]{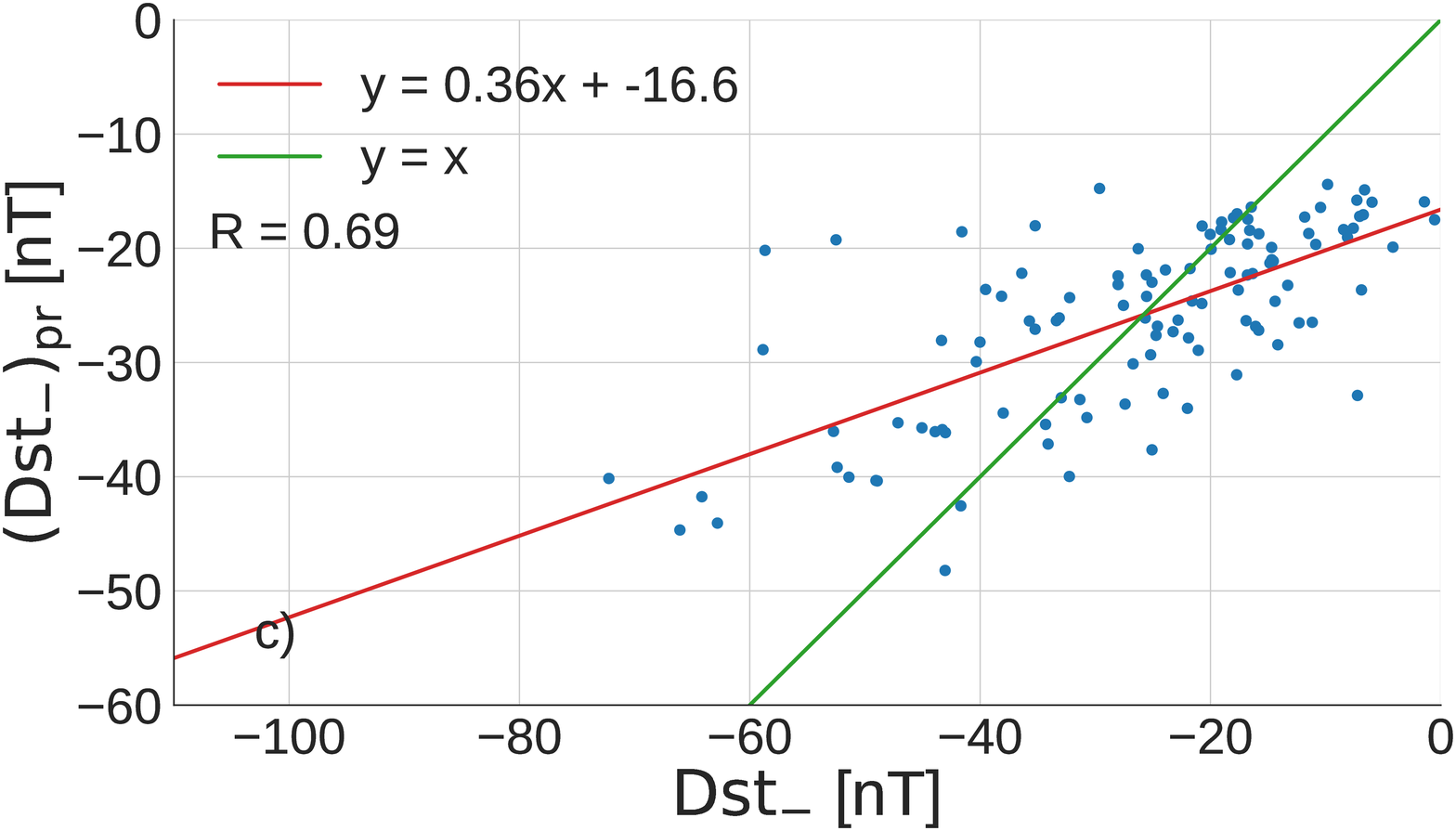}  
    \end{subfigure}
    \begin{subfigure}[b]{0.49\textwidth}
        \includegraphics[width=\linewidth]{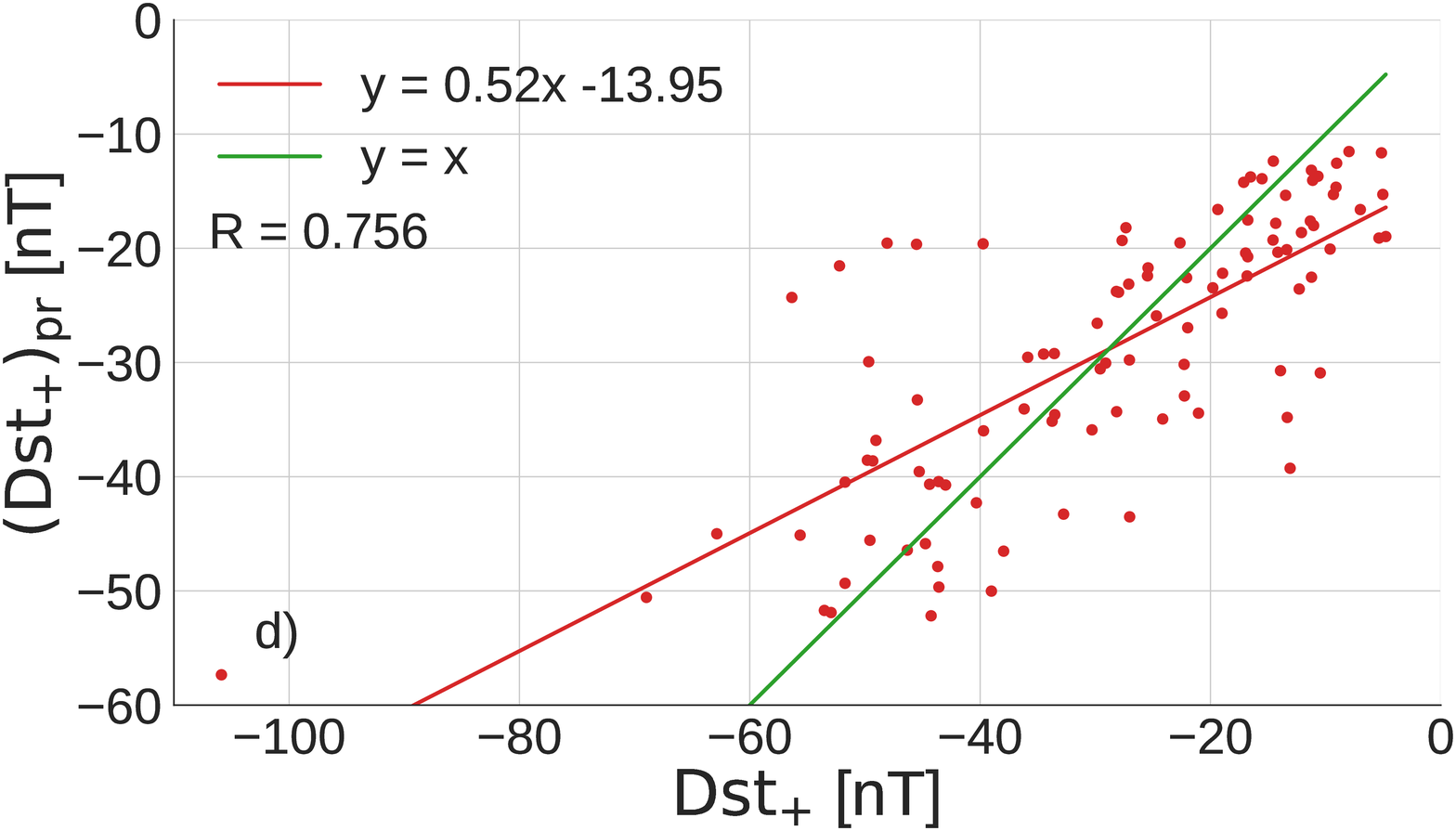}  
    \end{subfigure}
    \caption{
    The same as Figure~\ref{fig:Dstpred}, but for the actual HSS velocity measured at L1. 
    }
    \label{fig:true_Dstpred}
\end{figure*}

\begin{figure*}
    \centering
    \begin{subfigure}[b]{0.49\textwidth}
        \includegraphics[width=\linewidth]{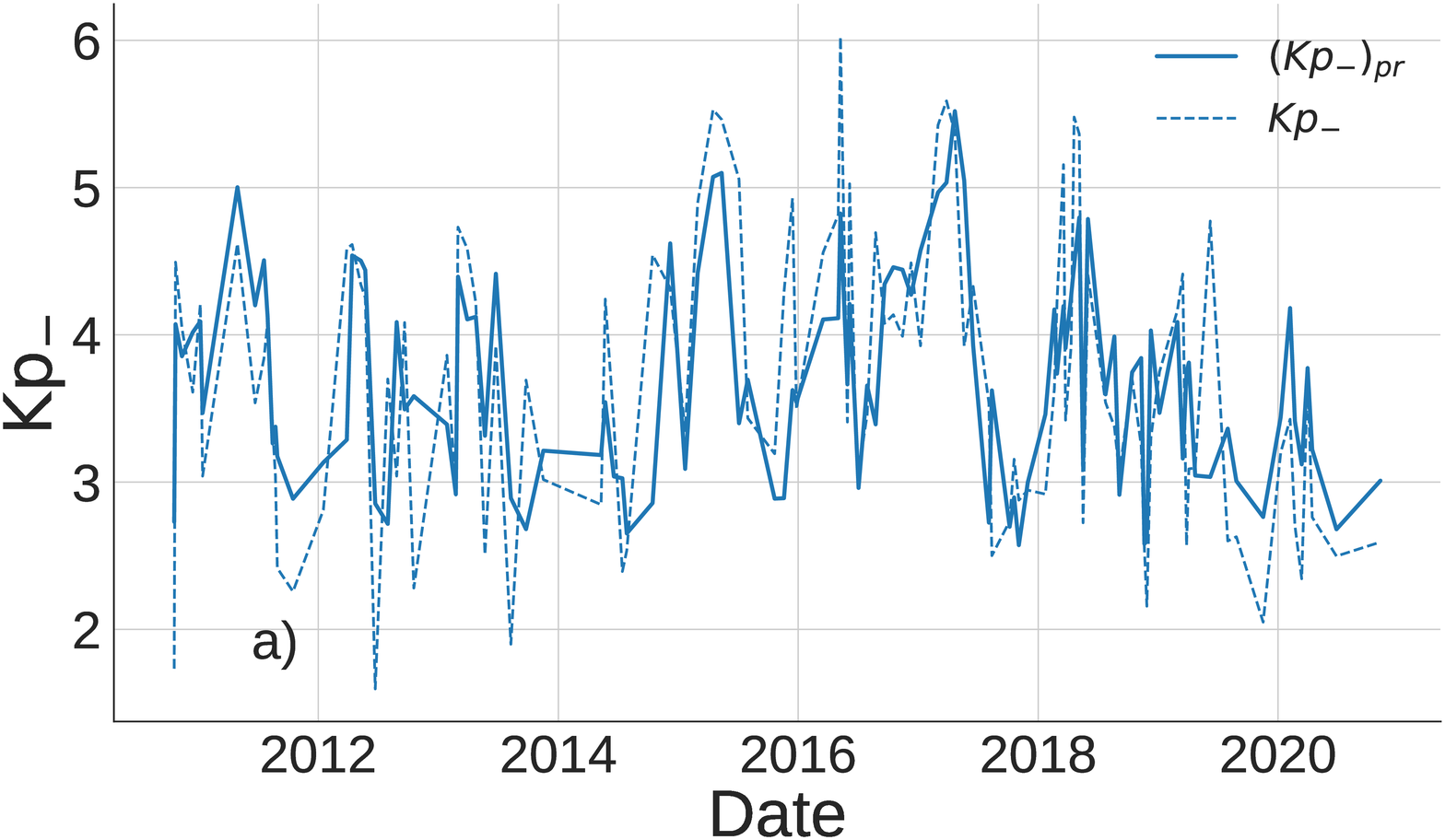}  
    \end{subfigure}
    \begin{subfigure}[b]{0.49\textwidth}
        \includegraphics[width=\linewidth]{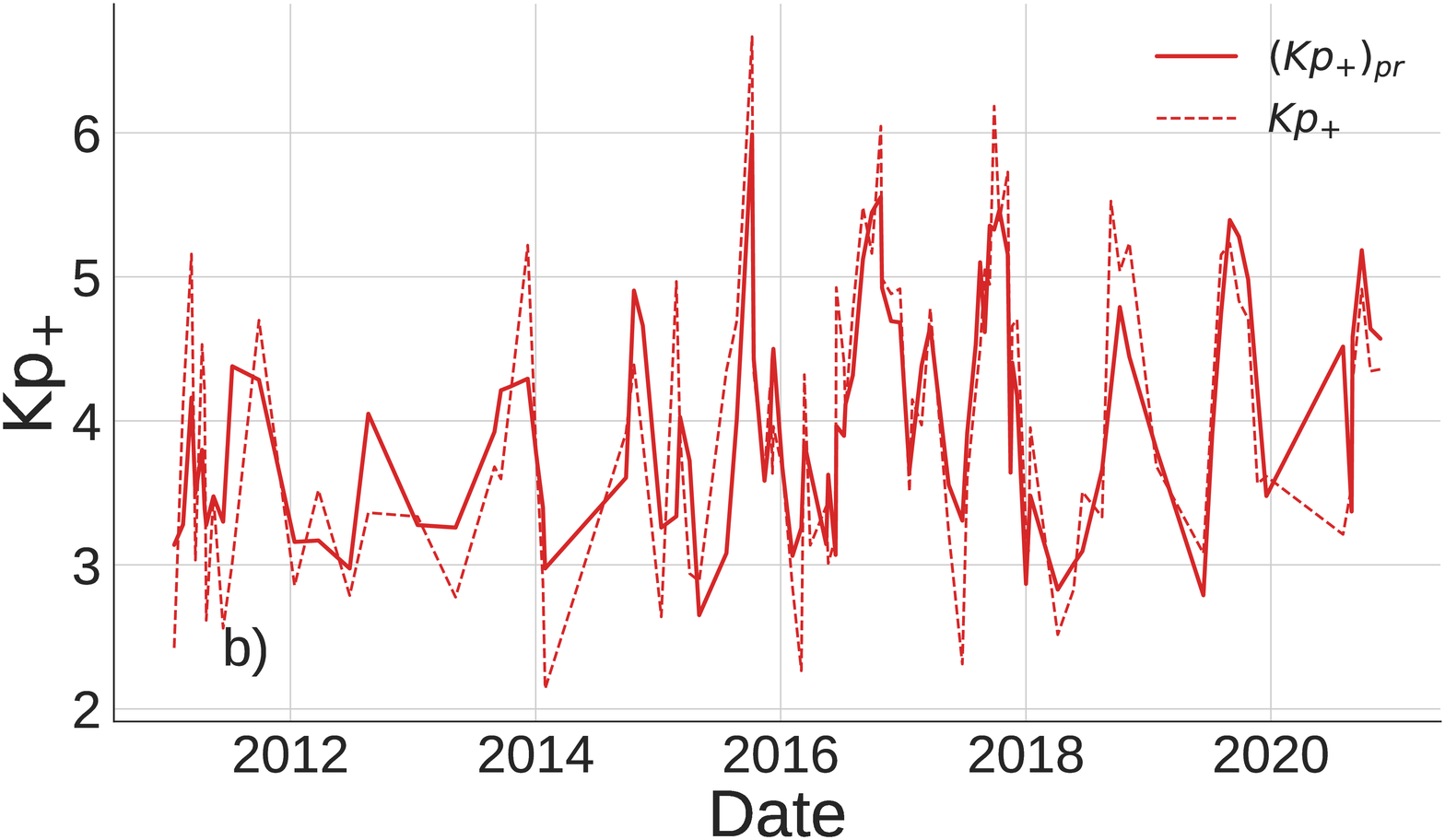}  
    \end{subfigure}
    
    \centering
    \begin{subfigure}[b]{0.49\textwidth}
        \includegraphics[width=\linewidth]{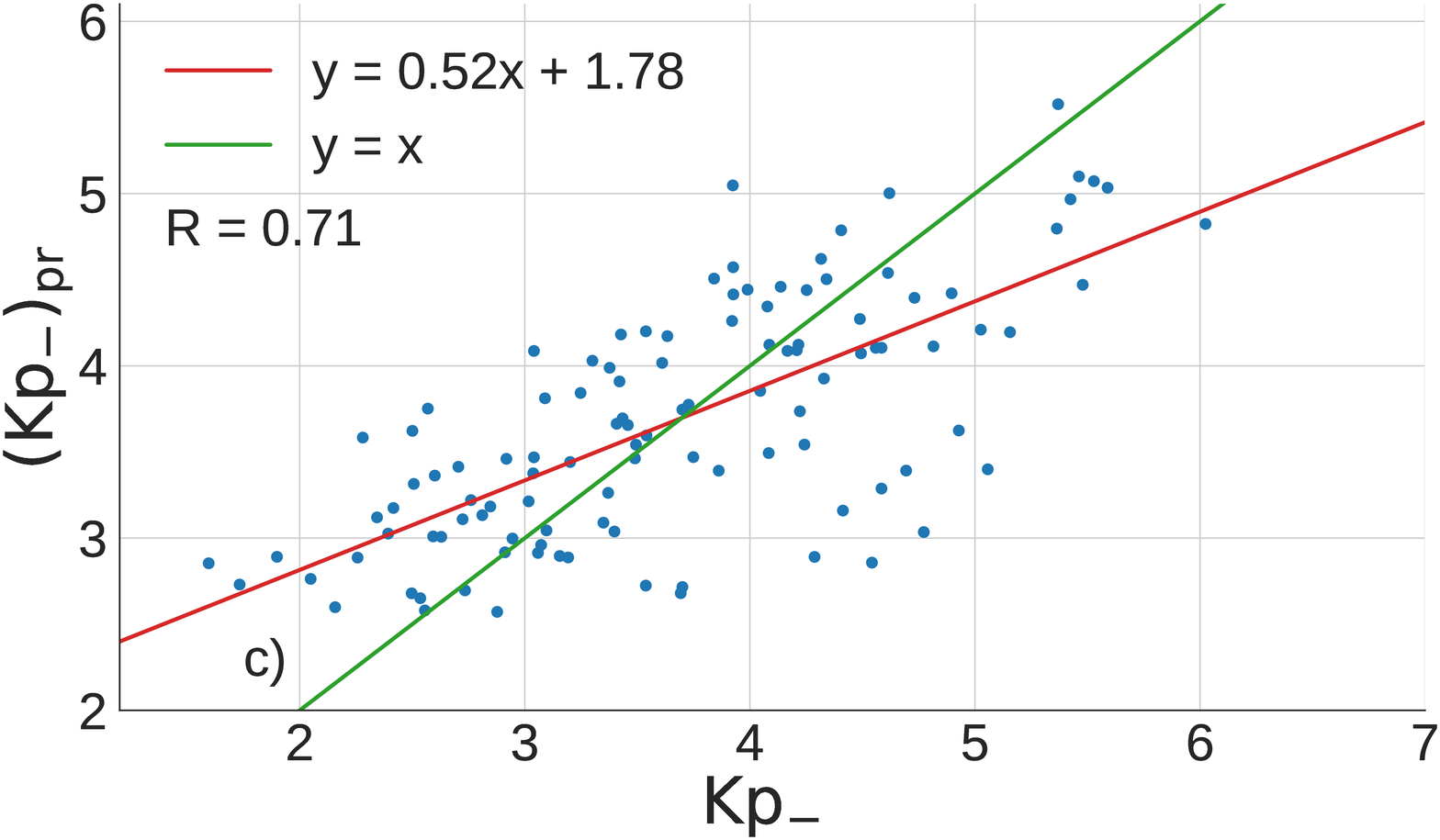}  
    \end{subfigure}
    \begin{subfigure}[b]{0.49\textwidth}
        \includegraphics[width=\linewidth]{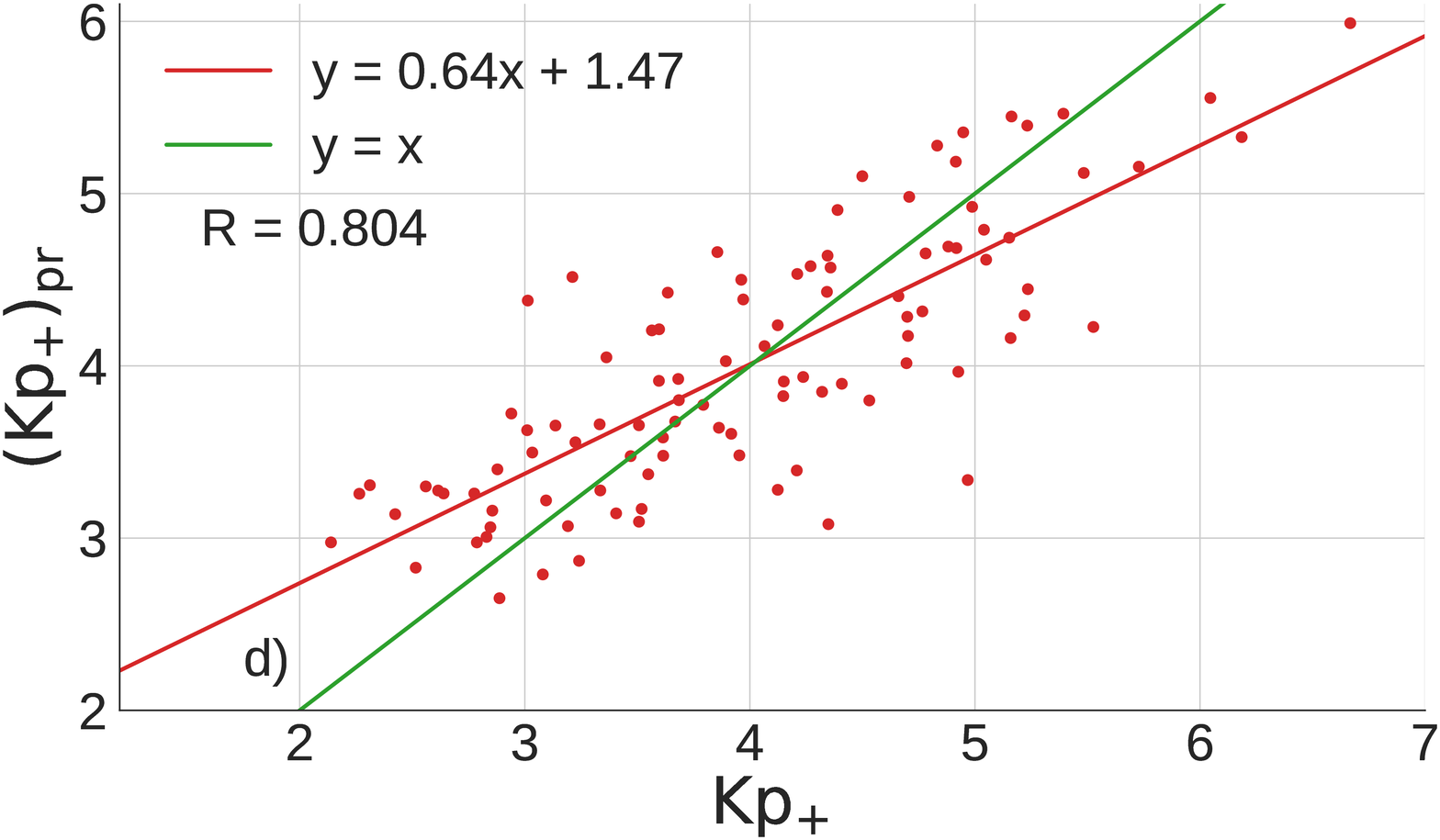}  
    \end{subfigure}
    \caption{
    The same as Figure~\ref{fig:Kppred}, but for the actual velocity measured at L1. 
    }
    \label{fig:true_Kppred}
\end{figure*}

To show that the major source of uncertainty in our model is due to our predictions of the solar wind speed, v$_\textrm{pr}$, we re-derive the forecasting of the geomagnetic indices using in Equation~\ref{eq:forecasting model} the actual solar wind velocity, v. 
Since it would reduce the lead warning period to 1--2 hours, this approach does not offer an alternative methodology for geomagnetic storm forecasting. However, it is a strategy to convey that more sophisticated solar wind speed forecastings are expected to significantly improve the final prediction of Dst and Kp. 
The results are shown in Figure~\ref{fig:true_Dstpred} and~\ref{fig:true_Kppred}, 
where we see that the local maxima and minima of Dst of Kp are now better approximated.  
In particular, panels (c) and (d) in each of the Figures show larger slopes of the linear least-square , i.e., closer to the ideal one-to-one correspondence. They increase from 0.3--0.4, when using the predicted velocities, to 0.5--0.6, when using the observed velocities.  
The correlation coefficients also enhance, from approximately 0.6--0.7 to 0.7--0.8.
In Table~\ref{tab:ForecastVer_dst_kp}, we see that when using the observed velocities overall the forecasting verification metrics improve. The largest improvement is observed for the Kp model with polarity away from the Sun. There, the MAE falls by \SI{19}{\percent}, from 0.57 to 0.46, and the RMSE decreases by a quarter, from 0.71 to 0.57.

\subsection{Model evaluation}

Finally, we perform a yearly evaluation of the Dst and Kp models. This analysis will show whether our model is biased by one or more years that outperform the others, and if a solar cycle trend exists.

For each year at hand, we leave out the events in the 5-year window centered on that year and fit a model on the remaining data set.
We then take a 3-year window centered on the year at hand as evaluation data set. The one-year standoff distance guarantees to prevent cross-talk, which could arise from long-lived CIR. Due to the decreased evaluation data set of only three years, we compute only one set of verification metrics that contains all events independent of their polarity to increase the statistical significance.

\begin{figure*}
    \centering
    \begin{subfigure}[b]{0.49\textwidth}
        \includegraphics[width=\linewidth]{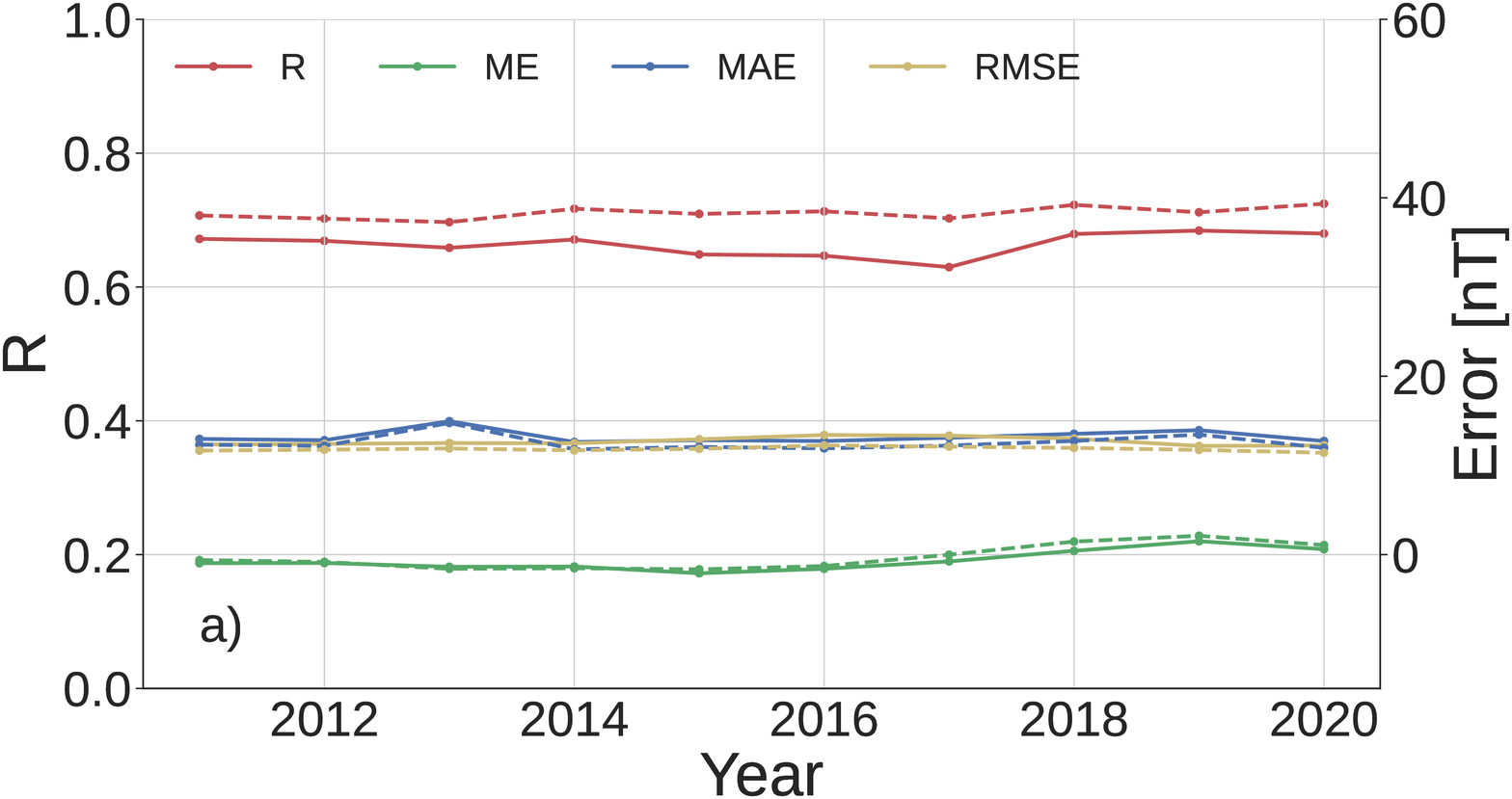}  
    \end{subfigure}
    \begin{subfigure}[b]{0.49\textwidth}
        \includegraphics[width=\linewidth]{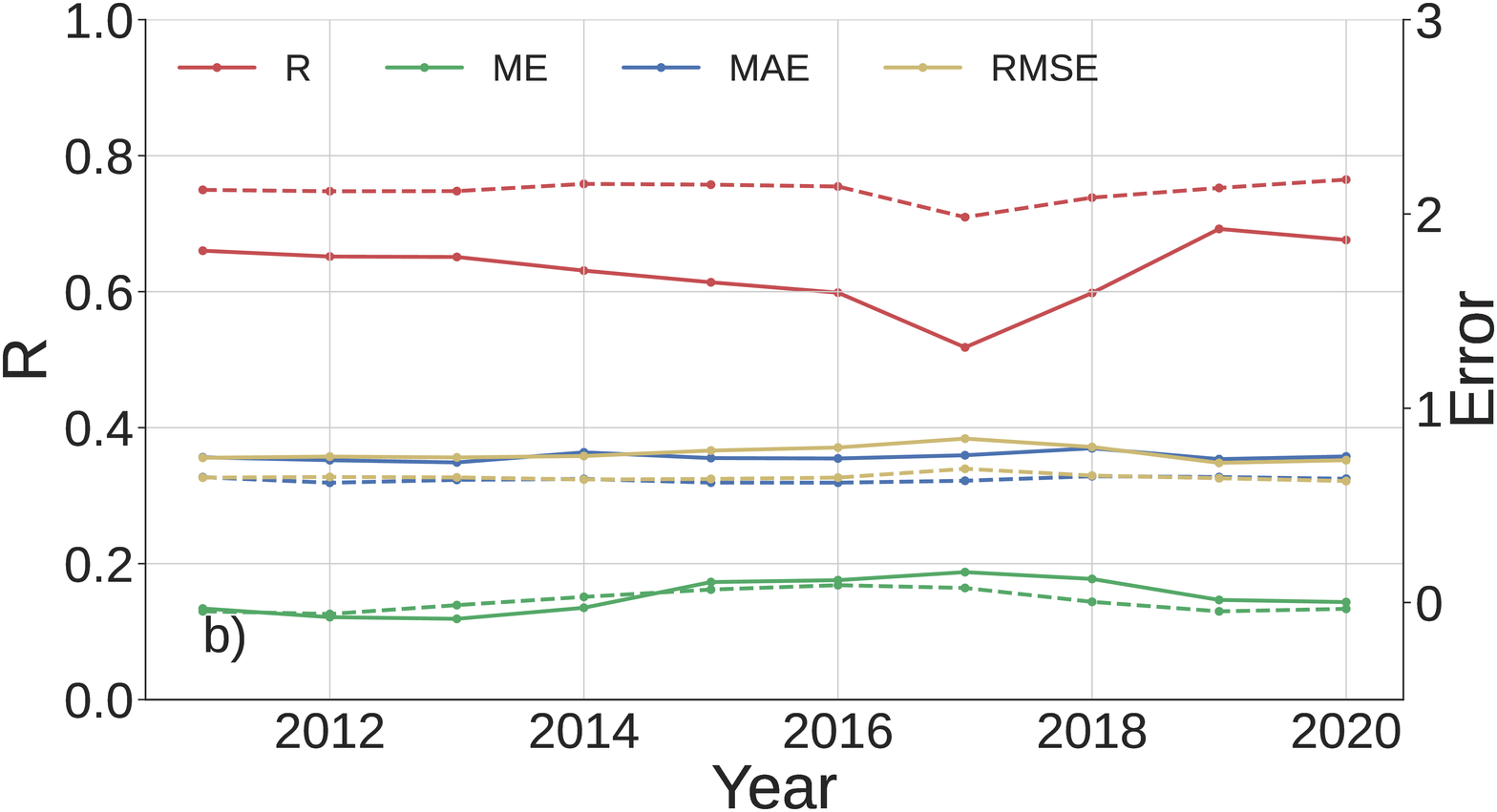}  
    \end{subfigure}
    \caption{ 
    Yearly evaluation of the Dst and Kp models. Panel (a) displays the forecasting verification metrics for a Dst model that employs predictions of the solar wind velocity (solid line), or the observed solar wind velocity (dashed line). Panel (b) is the same as panel (a), but for a Kp model.}
    \label{fig:evaluation model}
\end{figure*}

Fig. \ref{fig:evaluation model}(a) and \ref{fig:evaluation model}(b) show the metrics for the Dst and Kp index versus the center year of the evaluation dataset. In Fig. \ref{fig:evaluation model}(a), showing the metrics for the Dst index, we observe that all the metrics have a flat trend with only minor fluctuations. Hence, our forecast model works similarly well for each year, independent of the time in the solar cycle.
The metrics are close to those presented in Table \ref{tab:ForecastVer_dst_kp} for the Dst index. The correlation coefficient oscillates around R = 0.66, while in Table \ref{tab:ForecastVer_dst_kp} it is about 0.6--0.7 depending on the polarity of the CHs. This shows that we did not overfit our model.
In Fig. \ref{fig:evaluation model}(b), showing the evaluation for the Kp index, the results are similar to those of the Dst index, with a mean correlation coefficient of R = 0.63 compared to 0.65--0.67 in Table \ref{tab:ForecastVer_dst_kp}.

We then perform the same analysis using the observed solar wind velocity instead of the predicted one, which is shown by the dashed lines in Fig. \ref{fig:evaluation model}(a) and \ref{fig:evaluation model}(b). For the Dst index, we see that the correlation coefficient increases by a few percent, now being R = 0.71. The ME, MAE, and RMSE do not undergo noticeable changes.
For the Kp index, we find that all the metrics improve with R becoming $0.75$.
Moreover, while the Pearson coefficient computed with the observed solar wind velocity is nearly constant throughout the study period, the one computed with v$_\textrm{pr}$ has a sudden drop around 2017. This suggests a less good forecast of the solar wind velocity in that period, probably due to a long-lived CIR being an outlier in our dataset.

 \section{Conclusions}
\label{conclusion}
In this paper we investigated the possibility to predict the amplitudes of the CIR/HSS related Dst dips and Kp peaks by employing remote sensing observations of the Sun. We focused on the period 2010--2020, and 
found 256 geomagnetic events with their corresponding enhancement in the time-series of the coronal hole areas and coronal hole polarities derived from a 15 degree wide solar meridional slice, high-speed stream velocities and interplanetary magnetic field polarities. 
Amongst them, around $\SI{83}{\percent}$ of cases showed compatibility between the polarity of the IMF in HSS at L1 and the polarity of the open magnetic flux within coronal holes derived from the solar central-meridian slice. In this analysis, we only employed the events with matching polarity, corresponding to 212 geomagnetic events.

To build our forecasting model of Dst and Kp indices, we combined two steps. First, we fit the empirical relationship between solar wind velocity and coronal hole areas. 
The fit reveals a decent correlation between observed and predicted solar wind velocities, even though the latter were moderately underestimated.
Then, we assume Dst and Kp normalized to the solar wind velocity to be related to Bs. We take into account the Bs dependency on the polarity of coronal holes, by splitting up our data set into two subsets: periods where the polarity of the coronal holes was directed away from the Sun, and periods where it was directed toward the Sun. Moreover, we consider Bs depending on the orientation between the Earth's magnetic field axis and the IMF, by fitting Dst/v and Kp/v on the DOY.  
We find that the Dst/v and Kp/v timelines versus the day of the year resemble the \textit{pair of spectacles} pattern of the experimental Bs described by \cite{Verbanac2021}. They showed that Bs ordered according to polarity is maximized around the spring/fall equinox and does not vanish in fall/spring when the magnetic fields within the coronal holes in the central meridian point toward/away from the Sun.

To predict the peaks and dips of the geomagnetic indices, we multiplied the forecasted values of Dst/v and Kp/v, which are dependent on the day of the year, with the HSSs velocity predicted by the fractional area. The correlation coefficient between the observed and predicted Dst index is R $=0.63/0.73$ for high-speed streams having a polarity towards/away from the Sun. The corresponding correlation coefficients for the Kp index are R $= 0.65/0.67$. To show that a major source of uncertainty in our forecasting model is due to the scattered predictions of the solar wind speed from the coronal hole areas, we multiplied the forecasted values of Dst/v and Kp/v with high-speed streams velocity measured in situ at L1. We observe that the correlation coefficients and the forecasting verification metrics all undergo a significant improvement, from approximately 0.6--0.7 to 0.7--0.8. Furthermore, we performed an evaluation of the Dst and Kp models on a yearly basis. We observe that the correlation coefficients are quite stable and do not show any significant solar cycle trend.

Our findings show that, in general, the proposed technique allows for the prediction of HSS/CIR-driven geomagnetic storms directly from solar observations. This extends the lead time from hours, when one employs in situ solar wind data taken at Lagrangian point L1, to days, when employing solar observations. This increased warning time is critical for warnings of space weather conditions in the near-Earth environment and other space weather applications.

\section*{Acknowledgements}

S. J. Hofmeister acknowledges support by the DFG grant 448336908.
The authors are grateful to the NASA/SDO and the AIA, and HMI science teams for providing SDO data, National Space Science Data Center for the OMNI 2 database, to the National Oceanic and Atmospheric Administration (NOAA) for ACE solar wind data, to the WDC for Geomagnetism (Kyoto) for the Dst index data. The Kp index data are taken from the OMNI 2 database. 

\section*{Data Availability}
Data from SDO/AIA, SDO/HMI, ACE/SWEPAM, ACE/MAG, Wind/SWE, WDC-2, and the GSFC/SPDF OMNIWeb have open data policies.
Coronal hole data will be shared on reasonable request to the corresponding author.




\begin{thebibliography}{}
\makeatletter
\relax
\def\mn@urlcharsother{\let\do\@makeother \do\$\do\&\do\#\do\^\do\_\do\%\do\~}
\def\mn@doi{\begingroup\mn@urlcharsother \@ifnextchar [ {\mn@doi@}
  {\mn@doi@[]}}
\def\mn@doi@[#1]#2{\def\@tempa{#1}\ifx\@tempa\@empty \href
  {http://dx.doi.org/#2} {doi:#2}\else \href {http://dx.doi.org/#2} {#1}\fi
  \endgroup}
\def\mn@eprint#1#2{\mn@eprint@#1:#2::\@nil}
\def\mn@eprint@arXiv#1{\href {http://arxiv.org/abs/#1} {{\tt arXiv:#1}}}
\def\mn@eprint@dblp#1{\href {http://dblp.uni-trier.de/rec/bibtex/#1.xml}
  {dblp:#1}}
\def\mn@eprint@#1:#2:#3:#4\@nil{\def\@tempa {#1}\def\@tempb {#2}\def\@tempc
  {#3}\ifx \@tempc \@empty \let \@tempc \@tempb \let \@tempb \@tempa \fi \ifx
  \@tempb \@empty \def\@tempb {arXiv}\fi \@ifundefined
  {mn@eprint@\@tempb}{\@tempb:\@tempc}{\expandafter \expandafter \csname
  mn@eprint@\@tempb\endcsname \expandafter{\@tempc}}}

\bibitem[\protect\citeauthoryear{{Akasofu}}{{Akasofu}}{1981}]{Akasofu1981}
{Akasofu} S.~I.,  1981, \mn@doi [\planss] {10.1016/0032-0633(81)90121-5}, \href
  {https://ui.adsabs.harvard.edu/abs/1981P&SS...29.1151A} {29, 1151}

\bibitem[\protect\citeauthoryear{{Amata}, {Pallocchia}, {Consolini}, {Marcucci}
   \& {Bertello}}{{Amata} et~al.}{2008}]{Amata2008}
{Amata} E.,  {Pallocchia} G.,  {Consolini} G.,  {Marcucci} M.~F.,   {Bertello}
  I.,  2008, \mn@doi [Journal of Atmospheric and Solar-Terrestrial Physics]
  {10.1016/j.jastp.2007.08.041}, \href
  {https://ui.adsabs.harvard.edu/abs/2008JASTP..70..496A} {70, 496}

\bibitem[\protect\citeauthoryear{{Andriyas} \& {Andriyas}}{{Andriyas} \&
  {Andriyas}}{2015}]{Andriyas2015}
{Andriyas} T.,  {Andriyas} S.,  2015, \mn@doi [Journal of Atmospheric and
  Solar-Terrestrial Physics] {10.1016/j.jastp.2015.02.005}, \href
  {https://ui.adsabs.harvard.edu/abs/2015JASTP.125...10A} {125, 10}

\bibitem[\protect\citeauthoryear{{Bala} \& {Reiff}}{{Bala} \&
  {Reiff}}{2012}]{Bala2012}
{Bala} R.,  {Reiff} P.,  2012, \mn@doi [Space Weather] {10.1029/2012SW000779},
  \href {https://ui.adsabs.harvard.edu/abs/2012SpWea..10.6001B} {10, S06001}

\bibitem[\protect\citeauthoryear{{Brown}}{{Brown}}{1963}]{Brown1963}
{Brown} R.~G.,  1963, {Smoothing forecasting and prediction in discrete time
  series}.
Prentice Hall, New Jersey

\bibitem[\protect\citeauthoryear{{Burton}, {McPherron}  \& {Russell}}{{Burton}
  et~al.}{1975}]{Burton1975}
{Burton} R.~K.,  {McPherron} R.~L.,   {Russell} C.~T.,  1975, \mn@doi [\jgr]
  {10.1029/JA080i031p04204}, \href
  {https://ui.adsabs.harvard.edu/abs/1975JGR....80.4204B} {80, 4204}

\bibitem[\protect\citeauthoryear{{Cane} \& {Richardson}}{{Cane} \&
  {Richardson}}{2003}]{RichardsonCane2003}
{Cane} H.~V.,  {Richardson} I.~G.,  2003, \mn@doi [Journal of Geophysical
  Research (Space Physics)] {10.1029/2002JA009817}, \href
  {https://ui.adsabs.harvard.edu/abs/2003JGRA..108.1156C} {108, 1156}

\bibitem[\protect\citeauthoryear{{Carovillano} \& {Siscoe}}{{Carovillano} \&
  {Siscoe}}{1969}]{Carovillano1969}
{Carovillano} R.~L.,  {Siscoe} G.~L.,  1969, \mn@doi [\solphys]
  {10.1007/BF00155388}, \href
  {https://ui.adsabs.harvard.edu/abs/1969SoPh....8..401C} {8, 401}

\bibitem[\protect\citeauthoryear{Chilès \& Delfiner}{Chilès \&
  Delfiner}{2012}]{Chiles2012}
Chilès J.-P.,  Delfiner P.,  2012, \mn@doi [Wiley Series In Probability and
  Statistics] {10.1002/9781118136188}

\bibitem[\protect\citeauthoryear{{Choi}, {Moon}, {Choi}, {Baek}, {Kim}, {Cho}
  \& {Choe}}{{Choi} et~al.}{2009}]{Choi2009}
{Choi} Y.,  {Moon} Y.~J.,  {Choi} S.,  {Baek} J.-H.,  {Kim} S.~S.,  {Cho}
  K.~S.,   {Choe} G.~S.,  2009, \mn@doi [\solphys] {10.1007/s11207-008-9296-3},
  \href {https://ui.adsabs.harvard.edu/abs/2009SoPh..254..311C} {254, 311}

\bibitem[\protect\citeauthoryear{{Cranmer}}{{Cranmer}}{2009}]{Cranmer2009}
{Cranmer} S.~R.,  2009, \mn@doi [Living Reviews in Solar Physics]
  {10.12942/lrsp-2009-3}, \href
  {https://ui.adsabs.harvard.edu/abs/2009LRSP....6....3C} {6, 3}

\bibitem[\protect\citeauthoryear{Echer, Tsurutani  \& Gonzalez}{Echer
  et~al.}{2013}]{Echer2013}
Echer E.,  Tsurutani B.~T.,   Gonzalez W.~D.,  2013, Journal of Geophysical
  Research: Space Physics, 118, 385

\bibitem[\protect\citeauthoryear{{Gonzalez} \& {Echer}}{{Gonzalez} \&
  {Echer}}{2005}]{Gonzalez2005}
{Gonzalez} W.~D.,  {Echer} E.,  2005, \mn@doi [\grl] {10.1029/2005GL023486},
  \href {https://ui.adsabs.harvard.edu/abs/2005GeoRL..3218103G} {32, L18103}

\bibitem[\protect\citeauthoryear{{Gosling}, {Hundhausen}, {Pizzo}  \&
  {Asbridge}}{{Gosling} et~al.}{1972}]{Gosling1972}
{Gosling} J.~T.,  {Hundhausen} A.~J.,  {Pizzo} V.,   {Asbridge} J.~R.,  1972,
  \mn@doi [\jgr] {10.1029/JA077i028p05442}, \href
  {https://ui.adsabs.harvard.edu/abs/1972JGR....77.5442G} {77, 5442}

\bibitem[\protect\citeauthoryear{{Hofmeister}, {Veronig}, {Reiss}, {Temmer},
  {Vennerstrom}, {Vr{\v{s}}nak}  \& {Heber}}{{Hofmeister}
  et~al.}{2017}]{Hofmeister2017}
{Hofmeister} S.~J.,  {Veronig} A.,  {Reiss} M.~A.,  {Temmer} M.,  {Vennerstrom}
  S.,  {Vr{\v{s}}nak} B.,   {Heber} B.,  2017, \mn@doi [\apj]
  {10.3847/1538-4357/835/2/268}, \href
  {https://ui.adsabs.harvard.edu/abs/2017ApJ...835..268H} {835, 268}

\bibitem[\protect\citeauthoryear{{Hofmeister}, {Veronig}, {Temmer},
  {Vennerstrom}, {Heber}  \& {Vr{\v s}nak}}{{Hofmeister}
  et~al.}{2018}]{Hofmeister2018}
{Hofmeister} S.~J.,  {Veronig} A.,  {Temmer} M.,  {Vennerstrom} S.,  {Heber}
  B.,   {Vr{\v s}nak} B.,  2018, \mn@doi [J. Geophys. Res. (Space Physics)]
  {10.1002/2017JA024586}, \href
  {http://adsabs.harvard.edu/abs/2018JGRA..123.1738H} {123, 1738}

\bibitem[\protect\citeauthoryear{{Hofmeister}, {Veronig}, {Poedts}, {Samara}
  \& {Magdalenic}}{{Hofmeister} et~al.}{2020}]{Hofmeister2020}
{Hofmeister} S.~J.,  {Veronig} A.~M.,  {Poedts} S.,  {Samara} E.,
  {Magdalenic} J.,  2020, \mn@doi [\apjl] {10.3847/2041-8213/ab9d19}, \href
  {https://ui.adsabs.harvard.edu/abs/2020ApJ...897L..17H} {897, L17}

\bibitem[\protect\citeauthoryear{{Hofmeister} et~al.,}{{Hofmeister}
  et~al.}{2022}]{Hofmeister2022}
{Hofmeister} S.~J.,  et~al., 2022, \mn@doi [\aap]
  {10.1051/0004-6361/202141919}, \href
  {https://ui.adsabs.harvard.edu/abs/2022A&A...659A.190H} {659, A190}

\bibitem[\protect\citeauthoryear{{Kane} \& {Echer}}{{Kane} \&
  {Echer}}{2007}]{Kane2007}
{Kane} R.~P.,  {Echer} E.,  2007, \mn@doi [Journal of Atmospheric and
  Solar-Terrestrial Physics] {10.1016/j.jastp.2007.03.008}, \href
  {https://ui.adsabs.harvard.edu/abs/2007JASTP..69.1009K} {69, 1009}

\bibitem[\protect\citeauthoryear{{Katus}, {Liemohn}, {Ionides}, {Ilie},
  {Welling}  \& {Sarno-Smith}}{{Katus} et~al.}{2015}]{Katus2015}
{Katus} R.~M.,  {Liemohn} M.~W.,  {Ionides} E.~L.,  {Ilie} R.,  {Welling} D.,
  {Sarno-Smith} L.~K.,  2015, \mn@doi [Journal of Geophysical Research (Space
  Physics)] {10.1002/2014JA020712}, \href
  {https://ui.adsabs.harvard.edu/abs/2015JGRA..120..310K} {120, 310}

\bibitem[\protect\citeauthoryear{{Lemen} et~al.,}{{Lemen}
  et~al.}{2012}]{Lemen2012}
{Lemen} J.~R.,  et~al., 2012, \mn@doi [\solphys] {10.1007/s11207-011-9776-8},
  \href {https://ui.adsabs.harvard.edu/abs/2012SoPh..275...17L} {275, 17}

\bibitem[\protect\citeauthoryear{{Lepping} et~al.,}{{Lepping}
  et~al.}{1995}]{Lepping1995}
{Lepping} R.~P.,  et~al., 1995, \mn@doi [\ssr] {10.1007/BF00751330}, \href
  {https://ui.adsabs.harvard.edu/abs/1995SSRv...71..207L} {71, 207}

\bibitem[\protect\citeauthoryear{{Lundstedt}, {Gleisner}  \&
  {Wintoft}}{{Lundstedt} et~al.}{2002}]{Lundstedt2002}
{Lundstedt} H.,  {Gleisner} H.,   {Wintoft} P.,  2002, \mn@doi [\grl]
  {10.1029/2002GL016151}, \href
  {https://ui.adsabs.harvard.edu/abs/2002GeoRL..29.2181L} {29, 2181}

\bibitem[\protect\citeauthoryear{{Mansilla}}{{Mansilla}}{2008}]{Mansilla2008}
{Mansilla} G.~A.,  2008, \mn@doi [\physscr] {10.1088/0031-8949/78/04/045902},
  \href {https://ui.adsabs.harvard.edu/abs/2008PhyS...78d5902M} {78, 045902}

\bibitem[\protect\citeauthoryear{{McComas}, {Bame}, {Barker}, {Feldman},
  {Phillips}, {Riley}  \& {Griffee}}{{McComas} et~al.}{1998}]{McComas1998}
{McComas} D.~J.,  {Bame} S.~J.,  {Barker} P.,  {Feldman} W.~C.,  {Phillips}
  J.~L.,  {Riley} P.,   {Griffee} J.~W.,  1998, \mn@doi [\ssr]
  {10.1023/A:1005040232597}, \href
  {https://ui.adsabs.harvard.edu/abs/1998SSRv...86..563M} {86, 563}

\bibitem[\protect\citeauthoryear{{Neugebauer}, {Liewer}, {Smith}, {Skoug}  \&
  {Zurbuchen}}{{Neugebauer} et~al.}{2002}]{Neugebauer2002}
{Neugebauer} M.,  {Liewer} P.~C.,  {Smith} E.~J.,  {Skoug} R.~M.,   {Zurbuchen}
  T.~H.,  2002, \mn@doi [Journal of Geophysical Research (Space Physics)]
  {10.1029/2001JA000306}, \href
  {https://ui.adsabs.harvard.edu/abs/2002JGRA..107.1488N} {107, 1488}

\bibitem[\protect\citeauthoryear{{Nolte} et~al.,}{{Nolte}
  et~al.}{1976}]{Nolte1976}
{Nolte} J.~T.,  et~al., 1976, \mn@doi [\solphys] {10.1007/BF00149859}, \href
  {https://ui.adsabs.harvard.edu/abs/1976SoPh...46..303N} {46, 303}

\bibitem[\protect\citeauthoryear{{O'Brien} \& {McPherron}}{{O'Brien} \&
  {McPherron}}{2000a}]{OBrien2000a}
{O'Brien} T.~P.,  {McPherron} R.~L.,  2000a, \mn@doi [Journal of Atmospheric
  and Solar-Terrestrial Physics] {10.1016/S1364-6826(00)00072-9}, \href
  {https://ui.adsabs.harvard.edu/abs/2000JASTP..62.1295O} {62, 1295}

\bibitem[\protect\citeauthoryear{{O'Brien} \& {McPherron}}{{O'Brien} \&
  {McPherron}}{2000b}]{OBrien2000b}
{O'Brien} T.~P.,  {McPherron} R.~L.,  2000b, \mn@doi [\jgr]
  {10.1029/1998JA000437}, \href
  {https://ui.adsabs.harvard.edu/abs/2000JGR...105.7707O} {105, 7707}

\bibitem[\protect\citeauthoryear{Ogilvie \& Desch}{Ogilvie \&
  Desch}{1997}]{Ogilvie1997}
Ogilvie K.,  Desch M.,  1997, Advances in Space Research, 20, 559

\bibitem[\protect\citeauthoryear{{Ogilvie} et~al.,}{{Ogilvie}
  et~al.}{1995}]{Ogilvie1995}
{Ogilvie} K.~W.,  et~al., 1995, \mn@doi [\ssr] {10.1007/BF00751326}, \href
  {https://ui.adsabs.harvard.edu/abs/1995SSRv...71...55O} {71, 55}

\bibitem[\protect\citeauthoryear{{Owens}, {Arge}, {Spence}  \&
  {Pembroke}}{{Owens} et~al.}{2005}]{Owens2005}
{Owens} M.~J.,  {Arge} C.~N.,  {Spence} H.~E.,   {Pembroke} A.,  2005, \mn@doi
  [Journal of Geophysical Research (Space Physics)] {10.1029/2005JA011343},
  \href {https://ui.adsabs.harvard.edu/abs/2005JGRA..11012105O} {110, A12105}

\bibitem[\protect\citeauthoryear{{Owens} et~al.,}{{Owens}
  et~al.}{2008}]{Owens2008}
{Owens} M.~J.,  et~al., 2008, \mn@doi [Space Weather] {10.1029/2007SW000380},
  \href {https://ui.adsabs.harvard.edu/abs/2008SpWea...6.8001O} {6, S08001}

\bibitem[\protect\citeauthoryear{{Pallocchia}, {Amata}, {Consolini}, {Marcucci}
   \& {Bertello}}{{Pallocchia} et~al.}{2006}]{Pallocchia2006}
{Pallocchia} G.,  {Amata} E.,  {Consolini} G.,  {Marcucci} M.~F.,   {Bertello}
  I.,  2006, \mn@doi [Annales Geophysicae] {10.5194/angeo-24-989-2006}, \href
  {https://ui.adsabs.harvard.edu/abs/2006AnGeo..24..989P} {24, 989}

\bibitem[\protect\citeauthoryear{{Parker}}{{Parker}}{1958}]{Parker1958}
{Parker} E.~N.,  1958, \mn@doi [\apj] {10.1086/146579}, \href
  {https://ui.adsabs.harvard.edu/abs/1958ApJ...128..664P} {128, 664}

\bibitem[\protect\citeauthoryear{{Parker}}{{Parker}}{1963}]{Parker1963}
{Parker} E.~N.,  1963, {Interplanetary dynamical processes.}

\bibitem[\protect\citeauthoryear{{Pesnell}, {Thompson}  \&
  {Chamberlin}}{{Pesnell} et~al.}{2012}]{Pesnell2012}
{Pesnell} W.~D.,  {Thompson} B.~J.,   {Chamberlin} P.~C.,  2012, \mn@doi
  [\solphys] {10.1007/s11207-011-9841-3}, \href
  {https://ui.adsabs.harvard.edu/abs/2012SoPh..275....3P} {275, 3}

\bibitem[\protect\citeauthoryear{{Petrukovich} \& {Klimov}}{{Petrukovich} \&
  {Klimov}}{2000}]{Petrukovich2000}
{Petrukovich} A.~A.,  {Klimov} S.~I.,  2000, Cosmic Research, \href
  {https://ui.adsabs.harvard.edu/abs/2000CosRe..38..433P} {38, 433}

\bibitem[\protect\citeauthoryear{{Podladchikova} \&
  {Petrukovich}}{{Podladchikova} \&
  {Petrukovich}}{2012}]{Podladchikova2012storms}
{Podladchikova} T.~V.,  {Petrukovich} A.~A.,  2012, \mn@doi [Space Weather]
  {10.1029/2012SW000786}, \href
  {https://ui.adsabs.harvard.edu/abs/2012SpWea..10.7001P} {10, S07001}

\bibitem[\protect\citeauthoryear{{Podladchikova}, {Petrukovich}  \&
  {Yermolaev}}{{Podladchikova} et~al.}{2018}]{Podladchikova2018storms}
{Podladchikova} T.,  {Petrukovich} A.,   {Yermolaev} Y.,  2018, \mn@doi
  [Journal of Space Weather and Space Climate] {10.1051/swsc/2018017}, \href
  {https://ui.adsabs.harvard.edu/abs/2018JSWSC...8A..22P} {8, A22}

\bibitem[\protect\citeauthoryear{{Prölss}}{{Prölss}}{2004}]{Prolss2004}
{Prölss} G.~W.,  2004, {Physics of the Earth’s Space Environment}.
Springer Berlin Heidelberg, \mn@doi{10.1007/978-3-642-97123-5}

\bibitem[\protect\citeauthoryear{{Rasmussen} \& {Williams}}{{Rasmussen} \&
  {Williams}}{2005}]{Rasmussen2005}
{Rasmussen} C.~E.,  {Williams} C. K.~I.,  2005, {Gaussian processes for machine
  learning.}.
MIT Press

\bibitem[\protect\citeauthoryear{{Rast{\"a}tter} et~al.,}{{Rast{\"a}tter}
  et~al.}{2013}]{Rastatter2013}
{Rast{\"a}tter} L.,  et~al., 2013, \mn@doi [Space Weather] {10.1002/swe.20036},
  \href {https://ui.adsabs.harvard.edu/abs/2013SpWea..11..187R} {11, 187}

\bibitem[\protect\citeauthoryear{{Rathore}, {Dinesh}  \& {Parashar}}{{Rathore}
  et~al.}{2014}]{Rathore2014}
{Rathore} B.~S.,  {Dinesh} C.~G.,   {Parashar} K.~K.,  2014, Journal of Signal
  and Information Processing, 5, 42

\bibitem[\protect\citeauthoryear{{Reiss}, {Temmer}, {Veronig}, {Nikolic},
  {Vennerstrom}, {Sch{\"o}ngassner}  \& {Hofmeister}}{{Reiss}
  et~al.}{2016}]{Reiss2016}
{Reiss} M.~A.,  {Temmer} M.,  {Veronig} A.~M.,  {Nikolic} L.,  {Vennerstrom}
  S.,  {Sch{\"o}ngassner} F.,   {Hofmeister} S.~J.,  2016, \mn@doi [Space
  Weather] {10.1002/2016SW001390}, \href
  {https://ui.adsabs.harvard.edu/abs/2016SpWea..14..495R} {14, 495}

\bibitem[\protect\citeauthoryear{{Revallo}, {Valach}, {Hejda}  \&
  {Bochn{\'\i}{\v{c}}ek}}{{Revallo} et~al.}{2014}]{Revallo2014}
{Revallo} M.,  {Valach} F.,  {Hejda} P.,   {Bochn{\'\i}{\v{c}}ek} J.,  2014,
  \mn@doi [Journal of Atmospheric and Solar-Terrestrial Physics]
  {10.1016/j.jastp.2014.01.011}, \href
  {https://ui.adsabs.harvard.edu/abs/2014JASTP.110....9R} {110, 9}

\bibitem[\protect\citeauthoryear{{Robbins}, {Henney}  \& {Harvey}}{{Robbins}
  et~al.}{2006}]{Robbins2006}
{Robbins} S.,  {Henney} C.~J.,   {Harvey} J.~W.,  2006, \mn@doi [\solphys]
  {10.1007/s11207-006-0064-y}, \href
  {https://ui.adsabs.harvard.edu/abs/2006SoPh..233..265R} {233, 265}

\bibitem[\protect\citeauthoryear{{Rotter}, {Veronig}, {Temmer}  \&
  {Vr{\v{s}}nak}}{{Rotter} et~al.}{2012}]{Rotter2012}
{Rotter} T.,  {Veronig} A.~M.,  {Temmer} M.,   {Vr{\v{s}}nak} B.,  2012,
  \mn@doi [\solphys] {10.1007/s11207-012-0101-y}, \href
  {https://ui.adsabs.harvard.edu/abs/2012SoPh..281..793R} {281, 793}

\bibitem[\protect\citeauthoryear{{Rotter}, {Veronig}, {Temmer}  \&
  {Vr{\v{s}}nak}}{{Rotter} et~al.}{2015}]{Rotter2015}
{Rotter} T.,  {Veronig} A.~M.,  {Temmer} M.,   {Vr{\v{s}}nak} B.,  2015,
  \mn@doi [\solphys] {10.1007/s11207-015-0680-5}, \href
  {https://ui.adsabs.harvard.edu/abs/2015SoPh..290.1355R} {290, 1355}

\bibitem[\protect\citeauthoryear{{Russell} \& {McPherron}}{{Russell} \&
  {McPherron}}{1973}]{Russell1973}
{Russell} C.~T.,  {McPherron} R.~L.,  1973, \mn@doi [\jgr]
  {10.1029/JA078i001p00092}, \href
  {https://ui.adsabs.harvard.edu/abs/1973JGR....78...92R} {78, 92}

\bibitem[\protect\citeauthoryear{{Sarabhai}}{{Sarabhai}}{1963}]{Sarabhai1963}
{Sarabhai} V.,  1963, \mn@doi [\jgr] {10.1029/JZ068i005p01555}, \href
  {https://ui.adsabs.harvard.edu/abs/1963JGR....68.1555S} {68, 1555}

\bibitem[\protect\citeauthoryear{{Scherrer} et~al.,}{{Scherrer}
  et~al.}{2012}]{Scherrer2012}
{Scherrer} P.~H.,  et~al., 2012, \mn@doi [\solphys]
  {10.1007/s11207-011-9834-2}, \href
  {https://ui.adsabs.harvard.edu/abs/2012SoPh..275..207S} {275, 207}

\bibitem[\protect\citeauthoryear{{Sharifie}, {Lucas}  \& {Araabi}}{{Sharifie}
  et~al.}{2006}]{Sharifie2006}
{Sharifie} J.,  {Lucas} C.,   {Araabi} B.~N.,  2006, \mn@doi [Space Weather]
  {10.1029/2005SW000209}, \href
  {https://ui.adsabs.harvard.edu/abs/2006SpWea...4.6003S} {4, 06003}

\bibitem[\protect\citeauthoryear{{Shprits}, {Vasile}  \&
  {Zhelavskaya}}{{Shprits} et~al.}{2019}]{Shprits2019}
{Shprits} Y.~Y.,  {Vasile} R.,   {Zhelavskaya} I.~S.,  2019, \mn@doi [Space
  Weather] {10.1029/2018SW002141}, \href
  {https://ui.adsabs.harvard.edu/abs/2019SpWea..17.1219S} {17, 1219}

\bibitem[\protect\citeauthoryear{{Smith} \& {Wolfe}}{{Smith} \&
  {Wolfe}}{1976}]{Smith1976}
{Smith} E.~J.,  {Wolfe} J.~H.,  1976, \mn@doi [\grl] {10.1029/GL003i003p00137},
  \href {https://ui.adsabs.harvard.edu/abs/1976GeoRL...3..137S} {3, 137}

\bibitem[\protect\citeauthoryear{{Smith}, {L'Heureux}, {Ness}, {Acu{\~n}a},
  {Burlaga}  \& {Scheifele}}{{Smith} et~al.}{1998}]{Smith1998}
{Smith} C.~W.,  {L'Heureux} J.,  {Ness} N.~F.,  {Acu{\~n}a} M.~H.,  {Burlaga}
  L.~F.,   {Scheifele} J.,  1998, \mn@doi [\ssr] {10.1023/A:1005092216668},
  \href {https://ui.adsabs.harvard.edu/abs/1998SSRv...86..613S} {86, 613}

\bibitem[\protect\citeauthoryear{{Stone}, {Frandsen}, {Mewaldt}, {Christian},
  {Margolies}, {Ormes}  \& {Snow}}{{Stone} et~al.}{1998}]{Stone1998}
{Stone} E.~C.,  {Frandsen} A.~M.,  {Mewaldt} R.~A.,  {Christian} E.~R.,
  {Margolies} D.,  {Ormes} J.~F.,   {Snow} F.,  1998, \mn@doi [\ssr]
  {10.1023/A:1005082526237}, \href
  {https://ui.adsabs.harvard.edu/abs/1998SSRv...86....1S} {86, 1}

\bibitem[\protect\citeauthoryear{{Temerin} \& {Li}}{{Temerin} \&
  {Li}}{2002}]{Temerin2002}
{Temerin} M.,  {Li} X.,  2002, \mn@doi [Journal of Geophysical Research (Space
  Physics)] {10.1029/2001JA007532}, \href
  {https://ui.adsabs.harvard.edu/abs/2002JGRA..107.1472T} {107, 1472}

\bibitem[\protect\citeauthoryear{{Verbanac} \& {Bandi{\'c}}}{{Verbanac} \&
  {Bandi{\'c}}}{2021}]{Verbanac2021}
{Verbanac} G.,  {Bandi{\'c}} M.,  2021, \mn@doi [Solar Physics]
  {10.1007/s11207-021-01930-1}, \href
  {https://ui.adsabs.harvard.edu/abs/2021SoPh..296..183V} {296, 183}

\bibitem[\protect\citeauthoryear{{Verbanac}, {Vr{\v{s}}nak}, {Veronig}  \&
  {Temmer}}{{Verbanac} et~al.}{2011a}]{Verbanac2011a}
{Verbanac} G.,  {Vr{\v{s}}nak} B.,  {Veronig} A.,   {Temmer} M.,  2011a,
  \mn@doi [\aap] {10.1051/0004-6361/201014617}, \href
  {https://ui.adsabs.harvard.edu/abs/2011A&A...526A..20V} {526, A20}

\bibitem[\protect\citeauthoryear{{Verbanac}, {Vr{\v{s}}nak},
  {{\v{Z}}ivkovi{\'c}}, {Hojsak}, {Veronig}  \& {Temmer}}{{Verbanac}
  et~al.}{2011b}]{Verbanac2011b}
{Verbanac} G.,  {Vr{\v{s}}nak} B.,  {{\v{Z}}ivkovi{\'c}} S.,  {Hojsak} T.,
  {Veronig} A.~M.,   {Temmer} M.,  2011b, \mn@doi [\aap]
  {10.1051/0004-6361/201116615}, \href
  {https://ui.adsabs.harvard.edu/abs/2011A&A...533A..49V} {533, A49}

\bibitem[\protect\citeauthoryear{{Verbanac}, {{\v{Z}}ivkovi{\'c}},
  {Vr{\v{s}}nak}, {Bandi{\'c}}  \& {Hojsak}}{{Verbanac}
  et~al.}{2013}]{Verbanac2013}
{Verbanac} G.,  {{\v{Z}}ivkovi{\'c}} S.,  {Vr{\v{s}}nak} B.,  {Bandi{\'c}} M.,
   {Hojsak} T.,  2013, \mn@doi [\aap] {10.1051/0004-6361/201220417}, \href
  {https://ui.adsabs.harvard.edu/abs/2013A&A...558A..85V} {558, A85}

\bibitem[\protect\citeauthoryear{{Vr{\v{s}}nak}, {Temmer}  \&
  {Veronig}}{{Vr{\v{s}}nak} et~al.}{2007a}]{Vrsnak2007a}
{Vr{\v{s}}nak} B.,  {Temmer} M.,   {Veronig} A.~M.,  2007a, \mn@doi [\solphys]
  {10.1007/s11207-007-0285-8}, \href
  {https://ui.adsabs.harvard.edu/abs/2007SoPh..240..315V} {240, 315}

\bibitem[\protect\citeauthoryear{{Vr{\v{s}}nak}, {Temmer}  \&
  {Veronig}}{{Vr{\v{s}}nak} et~al.}{2007b}]{Vrsnak2007b}
{Vr{\v{s}}nak} B.,  {Temmer} M.,   {Veronig} A.~M.,  2007b, \mn@doi [\solphys]
  {10.1007/s11207-007-0311-x}, \href
  {https://ui.adsabs.harvard.edu/abs/2007SoPh..240..331V} {240, 331}

\bibitem[\protect\citeauthoryear{{Wei}, {Billings}  \& {Balikhin}}{{Wei}
  et~al.}{2004}]{Wei2004}
{Wei} H.~L.,  {Billings} S.~A.,   {Balikhin} M.,  2004, \mn@doi [Journal of
  Geophysical Research (Space Physics)] {10.1029/2003JA010332}, \href
  {https://ui.adsabs.harvard.edu/abs/2004JGRA..109.7212W} {109, A07212}

\bibitem[\protect\citeauthoryear{{Wing} et~al.,}{{Wing}
  et~al.}{2005}]{Wing2005}
{Wing} S.,  et~al., 2005, \mn@doi [Journal of Geophysical Research (Space
  Physics)] {10.1029/2004JA010500}, \href
  {https://ui.adsabs.harvard.edu/abs/2005JGRA..110.4203W} {110, A04203}

\bibitem[\protect\citeauthoryear{{Yermolaev}, {Nikolaeva}, {Lodkina}  \&
  {Yermolaev}}{{Yermolaev} et~al.}{2010}]{Yermolaev2010}
{Yermolaev} Y.~I.,  {Nikolaeva} N.~S.,  {Lodkina} I.~G.,   {Yermolaev} M.~Y.,
  2010, \mn@doi [Annales Geophysicae] {10.5194/angeo-28-2177-2010}, \href
  {https://ui.adsabs.harvard.edu/abs/2010AnGeo..28.2177Y} {28, 2177}

\bibitem[\protect\citeauthoryear{{Zhu}, {Billings}, {Balikhin}, {Wing}  \&
  {Coca}}{{Zhu} et~al.}{2006}]{Zhu2006}
{Zhu} D.,  {Billings} S.~A.,  {Balikhin} M.,  {Wing} S.,   {Coca} D.,  2006,
  \mn@doi [\grl] {10.1029/2005GL025022}, \href
  {https://ui.adsabs.harvard.edu/abs/2006GeoRL..33.4101Z} {33, L04101}

\makeatother
\end{thebibliography}



\appendix
\section{Gaussian Process Regression} \label{Sect:GPR}
    The Gaussian process regression \citep[GPR;][]{Rasmussen2005} is a nonparametric Bayesian approach for regression problems. 
    A Gaussian process (GP) itself can be thought as a generalization over functions of the Gaussian distribution. 
    In GPR, we assume the output $y$ of a function $f$ at input $x$ as $y = f(x)$. 
    The function $f$ can be completely specified by its mean and covariance functions, written as $f(x) \sim GP(m(x),\,K(x,x'))$, with $K$ computed using each pair ($x,x'$) in the input $x$.
    
    The first step of GPR is to approximate $f$ by another GP, $f^*\sim GP(m^*(x),\,K^*(x,x'))$, called prior, which specifies some properties of $f$ when noted.
    The prior mean function, $m^*(x)$, encodes the central tendency of the observed output, and it will be assumed to be zero. This choice is made for simplicity and requires the normalization of the observed output, so that the new mean is actually zero \citep[see][]{Chiles2012}. To normalize the observed data, we remove the mean and scale it to unit variance. Only when the solution of the GPR interpolation is available, we scale it back to the original representation.
    The  prior covariance function $K^*(x,x')$, also called kernel, encodes information about the shape and structure the function is expected to have.
    The most commonly used kernel in machine learning is the radial basis function (RBF). In this study, it is employed to estimate the solar wind velocity peaks as a function of CH area. It has a Gaussian form, defined as: 
    \begin{align}
        K^*_\textrm{RBF}(x,x') = \sigma_K^2\,\exp{\left(-\frac{(x-x')^2}{2l_K^2}\right)},
        \label{eq:rbf_kernel}
    \end{align}  
    where $x$ and $x'$ are every possible pair of CH area values.
    The hyperparameters $\sigma_K$ and $l_K$ are respectively the standard deviation of the output signal, which regulates the magnitude of the prior covariance at each point $x$,
    and the isotropic length-scale that controls the rate of decay of the covariance function between points $x$ and $x'$.
    To describe the ratios Dst/v and Kp/v as functions of the day of the year (DOY), we make use of a periodic kernel,
    \begin{align}
        K^*_\textrm{p}(x,x') = \sigma_K^2 \, \exp{\left(-\frac{2}{l_K^2}sin^2 \left(\pi \frac{|x-x'|}{p}\right)\right)},
        \label{eq:periodic_kernel}
    \end{align}
    where $x$ and $x'$ are every possible pair of DOY values.
    Here, $\sigma_K$ and $l_K$ are the same as in Equation~\ref{eq:rbf_kernel}, and $p$ is the distance between repetitions of the function, i.e. the period. In our model, $p$ is a fixed parameter equal to $365.24$ days, because it is meant to fit the annual periodicity of the geomagnetic activity.
    The kernel function we choose is used in the next step to get the covariance matrix $\Sigma$, such that the element on the \textit{i}-th row and \textit{j}-th column is $ \Sigma_{ij} = K^*(x_i,x_j)$, with $x_i$ and $x_j$ are two possible values assumed by $x$.
    
    Now let $y$ be the set of the known normalized output measurements of the observed inputs $X$, and $f$ the set of function values corresponding to the continuous set of inputs $x$. 
    According to the Bayesian inference, we can use the prior $f^*$ to find the predictive normalized GP of $f$:
    \begin{align}
    \begin{split}     
        & \textit{f}(x)|\,y \, \sim \, GP(\,m(x), \, K(x,x'))\\
        & m(x) = \, \Sigma(X,x)^T [\Sigma(X,X) + \sigma_\textrm{n}^2 I]^{-1}y\\
        & K(x,x') =  \, \Sigma(x,x')  + \sigma_\textrm{n}^2 I - \Sigma(X,x)^T \cdot \\ &
        \cdot[\Sigma(X,X) + \sigma_\textrm{n}^2 I]^{-1} \Sigma(X,x'))        .
        \label{eq:predictive_distribution}
    \end{split}
    \end{align}  
    Here $\sigma_\textrm{n}$ is the identically distributed noise in the observations, which results in a matrix with zeros everywhere except on the diagonal. This is represented by the product of the noise variance, $\sigma_\textrm{n}^2$, by the identity matrix $I$.
    
    Substituting in Equation~\ref{eq:predictive_distribution} the appropriate prior covariance, either Equation~\ref{eq:rbf_kernel} or \ref{eq:periodic_kernel} depending on the problem we are intending to solve, we get a set of hyperparameters, $\theta = \{\sigma_K,\,l_K,\,\sigma_\textrm{n} \}$.
    To find the optimal $\theta$, we use as loss function the negative log-likelihood, $L$, of the observed output $y$ given the observed input $X$, and we minimize it with respect to $\theta$. 
    \begin{align}
    \begin{split}
        L &= -\log(p(y|x,\,\theta)) =\\   
          &=\frac{1}{2}\log(\Sigma(X,X) + \sigma_\textrm{n}^2 I) + \frac{1}{2}y^T(\Sigma(X,X) + \sigma_\textrm{n}^2 I)^{-1} + \\ & +\frac{n}{2}\log(2\pi).
        \label{eq:loss function}
    \end{split}
    \end{align}
    When the optimal hyperparameters are substituted in Equation~\ref{eq:predictive_distribution}, $f$ is fully defined.  
    We can now write $f$ as $f(x) \sim GP(m(x),\sigma(x))$, where $\sigma$ is the standard deviation of the predictive function, computed as the squared root of the elements on the $K$ diagonal.
    Since we fitted the GPR on a set of normalized observed data, we obtained a normalized $f$. To scale it back, $m$ is multiplied by the standard deviation and summed to the mean both of the observed data, and $\sigma$ is multiplied by the standard deviation of the observed data.
    
\section{Forecast verification}\label{Sect:forecast verification}
    Forecast verification is applied to assess the quality of forecasts. Several scalar measures of forecast accuracy can be computed, such as the Pearson's correlation coefficient (R)
     \begin{align}
        R = \frac{\sum_{k=1}^n(f_k - \bar{f})(y_k-\bar{y})}{\sqrt{\sum_{k=1}^n(f_k - \bar{f})^2}\sqrt{\sum_{k=1}^n( y_k - \bar{y})^2}},
        \label{eq:R}
    \end{align} 
    where $f_k$ and $ y_k$ are the \textit{kth} element of $n$ total forecast and observation pairs, and $\bar{f}$ and $\bar{y}$ are their sample mean. R measures how close to a linear relation are two sets of data. Values closer to 1 represent stronger linear relationship, hence better predictions.
    
    The other scalar measures hereby used perform an error analysis of the predictions. The mean error (ME) is given by
     \begin{align}
        ME = \frac{1}{n} \sum_{k=1}^n (f_k - y_k),
        \label{eq:ME} 
    \end{align} 
    is the difference between the average forecast and the average observation.   
    
    The mean absolute error (MAE) is given by
     \begin{align}
        MAE = \frac{1}{n} \sum_{k=1}^n |f_k - y_k|,
        \label{eq:MAE}
    \end{align}  
    is the arithmetic mean of the absolute differences between the forecast and the observation pairs. Similar to the MAE is the root-mean-square
    error (RMSE),
     \begin{align}
        RMSE = \sqrt{\frac{1}{n} \sum_{k=1}^n (f_k - y_k)^2},
        \label{eq:RMSE}
    \end{align}
    which is the mean squared difference between forecast and observation value pairs. The RMSE is a typical magnitude for the forecast error being more sensitive to outliers. Note that ME, MAE and RMSE are equal to zero in the case that the forecast errors are equal to zero (that is $f_k = y_k$) and increase with increasing forecast errors.


\bsp	
\label{lastpage}
\end{document}